\documentclass[manuscript]{acmart}

\usepackage[ruled, vlined, linesnumbered]{algorithm2e}
\usepackage{subcaption}
\usepackage{multirow}
\usepackage{bm}
\usepackage{cleveref}
\usepackage[utf8]{inputenc}
\DeclareUnicodeCharacter{2212}{−}
\newcommand{\xhdr}[1]{\vspace{0.8mm}\noindent{{\bf #1}}}
\newcommand{\todo}[1]{\textcolor{red}{#1}}

\AtBeginDocument{%
  \providecommand\BibTeX{{%
    \normalfont B\kern-0.5em{\scshape i\kern-0.25em b}\kern-0.8em\TeX}}}

\setcopyright{acmcopyright}
\copyrightyear{2021}
\acmYear{2021}
\setcopyright{acmcopyright}\acmConference[RecSys '21]{Fifteenth ACM Conference on Recommender Systems}{September 27-October 1, 2021}{Amsterdam, Netherlands}
\acmBooktitle{Fifteenth ACM Conference on Recommender Systems (RecSys '21), September 27-October 1, 2021, Amsterdam, Netherlands}
\acmPrice{15.00}
\acmDOI{10.1145/3460231.3474275}
\acmISBN{978-1-4503-8458-2/21/09}

\begin{document}

\title{Black-Box Attacks on Sequential Recommenders via Data-Free Model Extraction}


\author{Zhenrui Yue}
\authornote{Both authors contributed equally to this research.}
\affiliation{%
  \institution{Technical University of Munich}
  \country{Germany}
}
\email{zhenrui.yue@tum.de}

\author{Zhankui He}
\authornotemark[1]
\affiliation{%
  \institution{University of California, San Diego}
  \country{USA}
}
\email{zhh004@ucsd.edu}

\author{Huimin Zeng}
\affiliation{%
  \institution{Technical University of Munich}
  \country{Germany}
}
\email{huimin.zeng@tum.de}

\author{Julian McAuley}
\affiliation{%
  \institution{University of California, San Diego}
  \country{USA}
}
\email{jmcauley@eng.ucsd.edu}

\renewcommand{\shortauthors}{Yue, et al.}

\begin{abstract}
%Can machine learning models weights be "stolen" and used to create attacks? 
We investigate 
%the question of 
whether \emph{model extraction} can be used to `steal' the weights of 
%victim 
sequential recommender systems,
%and threaten the victim with attack transferabilities.
and the potential threats posed to victims of such attacks.
This type of risk
has attracted attention in image and text classification, but to our knowledge not in recommender systems.
We argue that \emph{sequential} recommender systems are subject to unique vulnerabilities due to the specific autoregressive regimes used to train them.
%Recently, attentive sequential recommenders have been the trend for recent development in deep learning recommendation systems. 
Unlike many existing recommender attackers, which assume the dataset used to train the victim model is exposed to attackers,
%this paper considers 
we consider
a \emph{data-free} setting, where 
%real 
training data are not accessible.
Under this setting, we propose an API-based model extraction method via limited-budget synthetic data generation and knowledge distillation.  We investigate state-of-the-art models for sequential recommendation and show 
%how 
their vulnerability under model extraction and downstream attacks.

%We perform the attacks 
We perform attacks
in two stages. \emph{(1) Model extraction:} given different types of synthetic data and their labels retrieved from a black-box recommender, we extract the black-box model to a white-box model via distillation. 
% 0.688 agreement in top-10 recommendation, significantly better than 0.406 of baseline method. 
\emph{(2) Downstream attacks:} we attack the black-box model with adversarial samples generated by the white-box recommender. Experiments show the effectiveness of our data-free model extraction and downstream attacks on sequential recommenders in both \emph{profile pollution} and \emph{data poisoning} settings. 
\end{abstract}



% \keywords{datasets, neural networks, gaze detection, text tagging}

\maketitle

%%%%%%%%%%%%%%%%%%%%%%%%%%%%%%%%%%%%%%%%%%%%%%%%%%s

\section{Introduction}

Model extraction attacks~\cite{lowd2005adversarial, tramer2016stealing} try to make a local copy of 
%the serving 
a
machine learning model
%, only given the 
given only access to a
query API.
%it 
%Such attacks
Model extraction
%may lead to some issues like 
exposes issues such as
sensitive training information leakage~\cite{tramer2016stealing} and adversarial example attacks~\cite{papernot2017practical}. Recently, this topic 
%attracts great 
has attracted
attention in image classification~\cite{orekondy2019knockoff, papernot2017practical, zhou2020dast, kariyappa2020maze} and text classification~\cite{pal2019framework, Krishna2020Thieves}. In this work, we show that model extraction attacks
%could also be threats to 
also pose a threat to
sequential recommender systems.

Sequential models are a popular framework for personalized recommendation by capturing users' evolving interests and item-to-item transition patterns.
In recent years, various neural-network-based models, such as RNN and CNN frameworks
(e.g.~GRU4Rec\cite{hidasi2015session}, Caser\cite{tang2018personalized},
NARM\cite{li2017neural}) and Transformer frameworks (e.g.~SASRec\cite{kang2018self}, BERT4Rec\cite{sun2019bert4rec}) are widely used and consistently outperform non-sequential~\cite{rendle2012bpr, he2017neural} as well as traditional sequential models~\cite{rendle2010factorizing,he2016fusing}.
%have made significant improvements on accuracy and become increasingly popular. 
% emerge. 
% % without user profiling.
% Even supposing the model architecture is known, 
%to extract model weights from these sequential recommenders, some challenges should be considered: 
%However, most work aims to improve the accuracy of sequential recommenders or explore the applications on various scenarios, but the 
However, only a few works have studied attacks on recommenders, and have certain limitations:
%malicious attacks (and defense) are under-studied. 
%Current works have such limitations: 
(1) Few attack methods are tailored to sequential models. %Especially, though 
Attacks via adversarial machine learning 
%achieve the state-of-the-art in general recommendation attacks
have achieved the state-of-the-art in general recommendation settings~\cite{christakopoulou2019adversarial, fang2020influence, tang2020revisiting}, 
%their 
but
experiments are conducted on matrix-factorization models and 
%hard to be 
are hard to
directly apply to sequential recommendation; though some \emph{model-agnostic} attacks~\cite{lam2004shilling, burke2005limited} can be used in sequential settings, they 
%they 
heavily depend on heuristics and their effectiveness is often limited; (2) Many attack methods assume that full training data for the victim model is exposed to attackers~\cite{zhang2020practical, christakopoulou2019adversarial, fang2020influence, tang2020revisiting, li2016data}. 
%which usually 
Such data
can be used to train surrogate local models by attackers. 
%It is often 
However this setting is quite restrictive (or unrealistic), especially in 
%unrealistic because many recommenders are trained on 
implicit feedback settings (e.g. clicks, views), where data would be very difficult to obtain by an attacker.
%which is hard to access to due to privacy concerns.

%Therefore, 
We consider a \emph{data-free} setting, where no original training data is available to train a surrogate model. That is, 
%to say, 
we build a surrogate model without real training data but limited API queries.
%and 
We
first construct 
%the 
downstream attacks 
against
%to 
our surrogate (white-box) sequential recommender, then transfer the attacks to the victim (black-box) recommender. 

\begin{figure*}[t]
    \centering
    \includegraphics[width=0.8\linewidth]{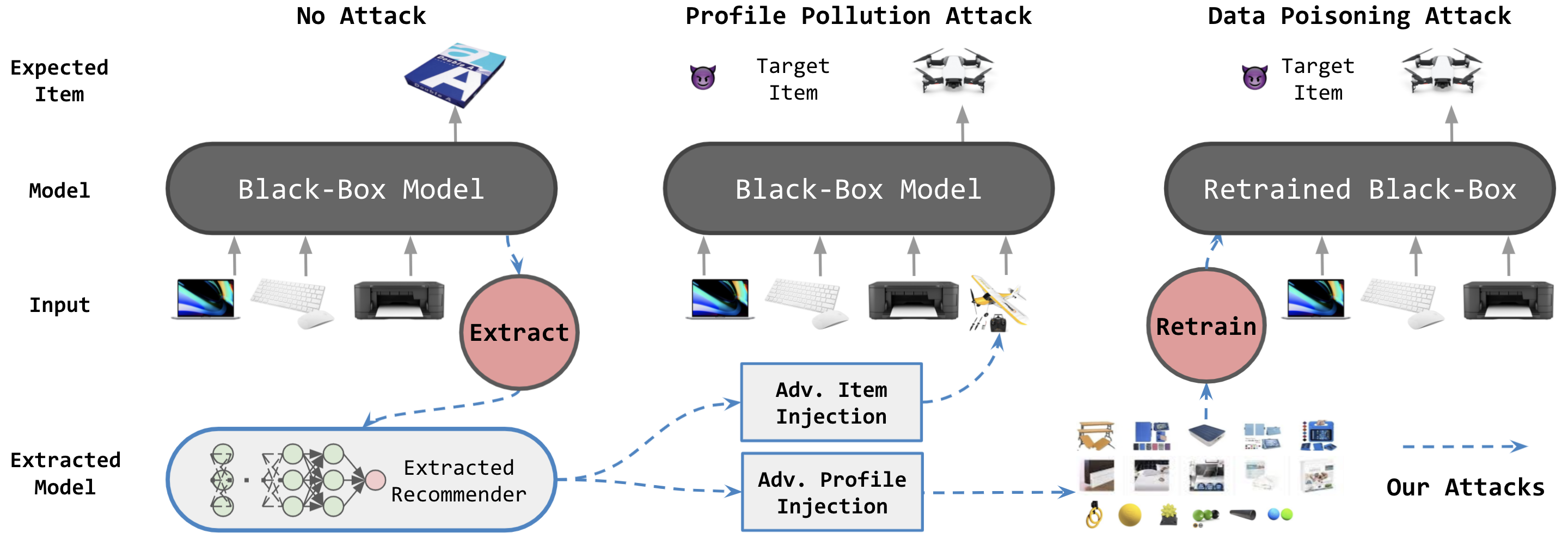}
    \caption{We illustrate two adversarial attack scenarios against sequential recommenders via model extraction.} 
    \vspace{-16pt}
    \label{fig:intro}
\end{figure*}

\emph{Model extraction} on sequential recommenders poses several challenges:
(1) no access to the orignial training dataset; (2) unlike image or text tasks, we cannot directly use surrogate datasets with semantic similarities; (3) APIs 
%only provide ranks of most likely items 
generally only provide rankings (rather than e.g.~probabilities)
and the query budget can be limited.
%of APIs query is limited. 
Considering
these challenges, sequential recommenders may seem relatively safe. However, 
%most 
noticing sequential recommenders are often trained in an autoregressive way 
%(i.e. based on the "next-item" prediction task), 
(i.e.,~predicting the next event in a sequence based on previous ones), our method shows the recommender itself can be used to generate sequential data which is similar to training data from the `real' data distribution. With this property and a sampling strategy: (1) `fake' training data can be constructed
%using which we show that  
that renders
sequential recommenders 
%are 
vulnerable 
%against 
to
model extraction; (2) the `fake' data from a limited number of API queries can resemble normal user behavior, which is difficult to detect.

\emph{Downstream attacks} are performed given the extracted surrogate model (see~\Cref{fig:intro}). But attack methods tailored to sequential recommenders are scarce~\cite{zhang2020practical}. In this work, we propose two attack methods with adversarial example techniques to current sequential recommenders, including \emph{profile pollution attacks} (that operate by `appending' items to users' logs) and \emph{data poisoning attacks} (where `fake' users 
%logs 
are generated
to bias the retrained model). 
%Moreover, 
% \add{Profile pollution performs user-specific attacks to promote target items with phishing and injection, where sensitive user profiles and activity information can be accessed using malwares~\cite{ren2015towards, lee2017all}}. Similarly, data poisoning is possible via creating fake users and achieve broader attacks by biasing the model towards target recommendation after retraining~\cite{yang2017fake}.
We extensively evaluate the effectiveness of our strategies in the setting that a black-box sequential model returns top-k ranked lists.

\section{Related Work}
\subsection{Model Extraction in Image and Text Tasks}

Model extraction attacks are proposed in~\cite{lowd2005adversarial, tramer2016stealing} by `stealing' model weights to make a local model copy%
%, and draw great attention in deep models
~\cite{orekondy2019knockoff, papernot2017practical, zhou2020dast, kariyappa2020maze, pal2019framework, Krishna2020Thieves}. Prior works are 
%mostly 
often
%based on 
concerned with
image classification. To extract the target model weights, \textit{JBDA}~\cite{orekondy2019knockoff} and \textit{KnockoffNets}~\cite{papernot2017practical} assume the attackers have access to partial training data or a surrogate dataset with semantic similarities. Recently, 
%some 
methods have been proposed in data-free settings. \textit{DaST}~\cite{zhou2020dast} adopted multi-branch Generative Adversarial Networks~\cite{goodfellow2014generative} to generate synthetic samples, which are subsequently labeled by the target model.
%subsequently. 
\textit{MAZE}~\cite{kariyappa2020maze} generated inputs that maximize the disagreement between the attacker and the target model. \textit{MAZE} used zeroth-order gradient estimation to optimize the generator module for accurate attacks. Because of the discrete nature of the input space, the methods above cannot transfer to sequential data directly. For Natural Language Processing (NLP) systems, 
% THIEVES~
\textit{THIEVES}~\cite{Krishna2020Thieves} studied model extraction attacks on BERT-based APIs~\cite{jacob2018bert}. 
%In this work, 
Even though attacks against BERT-based models exist for NLP,
where authors find \emph{random} word sequences and 
%some 
a surrogate dataset (e.g.~WikiText-103~\cite{merity2016pointer}) that can both create effective queries and retrieve labels to approximate
%copy the BERT model
%only slightly worse
%produce a BERT model close to 
%than 
the target model, we found that (1) in recommendation, it is hard to use surrogate datasets with semantic similarities (as is common in NLP); we also adopt \emph{random} item sequences as a baseline but their model extraction performance is limited. Therefore we generate data 
%with the 
following the
\emph{autoregressive} property of sequential recommenders; (2) Compared with NLP, it is harder to distill `ranking' (instead of classification) in recommendation. We design a pair-wise ranking loss to tackle the challenge; (3) downstream attacks after model extraction are under-explored, especially in recommendation, so our work also contributes in this regard.
%can be exploited 

\subsection{Attacks on Recommender Systems} 
\label{sec:related}
Existing works~\cite{yang2017fake, huang2021data} categorize attacks on recommender systems as \emph{profile pollution attacks} and \emph{data poisoning attacks}, affecting a recommender system in test and training phases respectively.

Profile pollution attacks aim to pollute a target user's profile (such as their view history)
%histories 
to manipulate the specific user's recommendation results.
%for this user. 
For example,~\cite{xing2013take} uses a cross-site request forgery (CSRF) technique~\cite{zeller2008cross}
to inject `fake' user views into target user logs in real-world websites including YouTube, Amazon and Google Search. However the strategy used in~\cite{xing2013take} to decide which `fake' item should be injected is a simple and heuristic without the knowledge of victim recommenders. In our work, given that we can extract victim recommender weights, more effective attacks can be investigated, such as \emph{evasion attacks} in general machine learning~\cite{goodfellow2014explaining,kurakin2016adversarial, papernot2017practical}. Note that in our work, we assume we can append items to target user logs with injection attacks via implanted malware~\cite{ren2015towards, lee2017all}, where item interactions could be added on users' behalf, therefore, we focus on injecting algorithm design and attack transferabilities (exact approaches for malware development and activity injection are cybersecurity tasks and beyond our research scope).

% (how to inject e.g.~clicks into user logs is a cybersecurity task and beyond our research scope).

Data poisoning attacks (\emph{a.k.a.} Shilling attacks~\cite{lam2004shilling, gunes2014shilling}) generate 
%well-designed items 
ratings from a number of fake users to poison training data. Some poisoning methods are recommender-agnostic~\cite{lam2004shilling, burke2005limited} (i.e.,~they do not consider the
%which do not consider the 
characteristics of the model architecture)
but they heavily depend on heuristics and often limit their effectiveness. 
%so the effectiveness is often limited.
Meanwhile, 
%many
poisoning attacks 
%are 
have been
proposed 
%to 
for
specific recommender architectures. For example,~\cite{li2016data, christakopoulou2019adversarial, tang2020revisiting} propose poisoning algorithms for matrix-factorization recommenders and~\cite{yang2017fake}
poisons co-visitation-based recommenders. Recently, as 
%with deep 
deep learning has been widely applied to recommendation,%
%learning widely applied in recommendation,
~\cite{huang2021data} investigate data poisoning in the neural collaborative filtering framework (NCF)~\cite{he2017neural}. The closest works to ours are perhaps LOKI~\cite{zhang2020practical}, because of the \emph{sequential} setting and~\cite{christakopoulou2019adversarial, fang2020influence, tang2020revisiting} 
which
adopt adversarial machine learning techniques to generate `fake' user profiles for matrix-factorization models. 
%LOKI is the first work to poison data in \emph{sequential} recommenders, utilizing reinforcement learning 
%algorithm 
%to train the attack agent. 
%However, one of the 
One
(fairly restrictive)
limitation of LOKI 
%is assuming 
is
the assumption that the
%that 
attacker can access
%full 
complete interaction data used for training. Another limitation is that LOKI is infeasible against deep networks (e.g. RNN / transformer), due to an unaffordable (even with tricks) Hessian computation.
%training dataset.
%in the recommendation system. 
%That is to say, 
%That the
%attacker 
%knows 
%could access 
%the
%full view logs of each user, which is 
%usually unrealistic. 
%such data is a fairly restrictive limitation.
In our work, black-box recommender weights are extracted for attacks without any real training data; we further design approximated and effective attacks to these recommenders.
%(i.e.,~data-free model extraction).

% \subsection{Adversarial Attacks}
%  The recent 
%  %prevalence of 
%  interest in adversarial examples \cite{szegedy2013intriguing} has 
%  %posed 
%  revealed threats in security-critical scenarios. Imperceptible perturbations added to inputs can cause deep neural models to output incorrect predictions. Powerful adversarial examples could be crafted using gradient-based methods \cite{goodfellow2014explaining, kurakin2016adversarial} or deep generative models \cite{odena2017conditional, wang2019gan}. Such adversarial examples can corrupt the performance of well-trained deep models dramatically. 
% %Moreover, in the context of machine learning, 

% Another form of adversarial attack is data poisoning \cite{barreno2006can,huang2011adversarial}. The most critical aspect of modern machine learning algorithms is the training data, typically collected from the real world. For a system trained on the training set, an adversarial attacker can inject synthesized examples which are deliberately biased; once this model is retrained on such a biased dataset, the resulting system 
% %is biased. 
% may carry the same bias.
 
\section{Framework}

 Our framework 
 %is two-stage: 
 has two stages:
 \emph{(1) Model extraction:} we generate informative synthetic data to train our white-box recommender that 
 %could 
 can
 rapidly close the gap between the victim recommender and ours via knowledge distillation~\cite{hinton2015distilling}; \emph{(2) Downstream attacks:} we propose gradient-based adversarial sample generation algorithms, which allows us to find effective adversarial sequences in the discrete item space from the \emph{white-box} recommender and achieve successful profile pollution or data poisoning attacks against the \emph{victim} recommender. 

\subsection{Setting}
To focus on model extraction and attacks against black-box sequential recommender systems, we first introduce some details regarding our setting. We formalize the problem with the following settings to define the research scope:

\begin{itemize}
    \item \emph{Unknown Weights:} Weights or metrics of the victim recommender are not provided.
    \item \emph{Data-Free:} Original training data is not available, and item statistics (e.g.~popularity) are not accessible.
    \item \emph{Limited API Queries:} Given some input data, the victim model API provides a ranked list of items (e.g.~top 100 recommended items). To avoid large numbers of API requests, we define budgets for the total number of victim model queries. Here, we treat each input sequence as one budget unit.
    \item \emph{(Partially) Known Architecture:} Although weights are confidential, model architectures are known (e.g.~we know the victim recommender is a transformer-based model). %We also investigate our methods with different white-box recommender architecture extracting the same the victim recommender, 
    We also relax this assumption to cases
    where the white-box recommender uses a different sequential model architecture from the victim recommender.
\end{itemize}

\subsection{Threat Model}

\subsubsection{Black-box Victim Recommender} Formally, we first denote $\bm{\mathcal{I}}$ as the discrete item space with $|\bm{\mathcal{I}}|$ elements. Given a sequence of length $T$ ordered by timestamp, i.e.,~$\bm{x} = [x_1,x_2,\ldots,x_T]$ where $x_i \in \bm{\mathcal{I}}$, a victim sequential recommender $\bm{f}_b$ is a \emph{black-box},
%because 
i.e.,~the weights are unknown. $\bm{f}_b$  should return a truncated ranked list over the next possible item in the item space $\bm{\mathcal{I}}$, i.e.,~$\hat{\bm{I}}^k = \bm{f}_b(\bm{x})$,
where $\hat{\bm{I}}^k$ is the truncated ranked list for the top-k recommended items. 
%In 
Many platforms % (e.g.~YouTube, Amazon)
%we can get access to these 
%ranked lists, such as YouTube and Amazon. 
surface ranked lists in such a form.

\subsubsection{White-box Surrogate Recommender} 
%For 
We construct
the white-box model $\bm{f}_w$
%, we construct it 
with two components: an embedding layer $\bm{f}_{w, e}$ and a sequential model $\bm{f}_{w, m}$ such that $\bm{f}_w(\bm{x}) = \bm{f}_{w, m}(\bm{f}_{w, e}(\bm{x}))$. Although the black-box model only returns a list of recommended items, the output scores $\hat{\bm{S}}_w$ of white-box model $\bm{f}_w$ over the input space $\bm{\mathcal{I}}$ 
%could 
can now
be accessed,
i.e.,~$\hat{\bm{S}}_w = \bm{f}_w(\bm{x})$.

\subsection{Attack Goal}

\subsubsection{Model Extraction}

As motivated previously, the first step is to extract a \textbf{white-box} model $\bm{f}_{w}$ from a trained, \textbf{black-box} victim model $\bm{f}_{b}$.  Without accessing the weights of $\bm{f}_{b}$, we obtain information from the black-box model by making limited 
%inquiries from it 
queries
and saving the predicted ranked list $\hat{\bm{I}}^k$
%for each inquiry. 
from each query.
In other words, we want to minimize the distance between the black-box $\bm{f}_{b}$ and white-box model $\bm{f}_{w}$ on 
%inquiry 
query
results. We are able to achieve the goal of learning a white-box model via knowledge distillation \cite{hinton2015distilling}. Mathematically, the model extraction process can be formulated as an optimization problem:
\begin{equation}
    \bm{f}^*_{w} = \arg\min_{\substack{\bm{f}_{w}}} \sum_{i = 1}^{|\bm{\mathcal{X}}|} \mathcal{L}_{\mathit{dis}}(\hat{\bm{I}}^{k^{(i)}}, \bm{f}_{w}(\bm{x}^{(i)})),
\end{equation}
where $\bm{\mathcal{X}}$ represents a set of sequences and $\bm{x}^{(i)} \in \bm{\mathcal{X}}$ with unique identifier $i$. We define $\bm{\mathcal{\hat{I}}}^k$ as a set of top-k predicted ranked lists, where $\hat{\bm{I}}^{k^{(i)}} \in \bm{\mathcal{\hat{I}}}^k$ is the black-box model output for $\bm{x}^{(i)}$. Note that data $(\bm{\mathcal{X}}, \mathcal{\hat{\bm{I}}}^k)$ are not real training data. Instead, they are generated with specific strategies, whose details will be included in~\Cref{sec:method}. $\mathcal{L}_{\mathit{dis}}$ is a loss function measuring the distance between two model outputs, such as a ranking loss.

\subsubsection{Downstream Attacks} We use the extracted white-box model $\bm{f}^*_w$ as the surrogate of the black-box model $\bm{f}_b$ to construct attacks. In this work, we investigate \textbf{targeted promotion attacks}, 
%namely to 
whose goal is to
increase the target item exposure to users as much as possible, which is 
%the most common case due to commercial interests
a common attack scenario~\cite{tang2020revisiting}. Note that targeted \emph{demotion} attacks can also be constructed with similar techniques. Formally, the objective of targeted promotion attacks is:

\begin{itemize}
    \item \emph{Profile Pollution Attack:} We define profile pollution attacks formally as the problem of finding the optimum injection items $\bm{z}^*$ 
    %\textbf{after} 
    (that should be items appended after
    the original sequence $\bm{x}$) that maximize the target item exposure $\bm{E}_{t}$, which can be characterized with common ranking measures like \emph{Recall} or \emph{NDCG}~\cite{kang2018self, sun2019bert4rec, li2017neural}:
        \begin{equation}
            \bm{z}^* = \arg\max_{\substack{\bm{z}} } \bm{E}_{t}(\bm{f}_{b}([\bm{x}; \bm{z}])),
        \end{equation}
    where $[\bm{x}; \bm{z}]$ refers to the concatenation of the sequence $\bm{x}$ and attacking items $\bm{z}$. Note that in profile pollution attacks setting, no retraining needed and this \textbf{user-specific profile $\bm{x}$ is assumed to be accessed and can be injected} (e.g.~using malwares~\cite{ren2015towards, lee2017all}; see~\Cref{sec:related}).
    
    \item \emph{Data Poisoning Attack:} Similarly, poisoning attacks 
    %could 
    can
    be 
    %seen as to find 
    viewed as finding
    biased injection profiles $\mathcal{\bm{Z}}$, such that after retraining, 
    the
    recommender 
    %carries the same 
    propagates the
    bias and is more likely to recommend the target. $\mathcal{\bm{Z}}\cup \mathcal{\bm{X}}$ refers to 
    the
    injection of fake profiles $\mathcal{\bm{Z}}$ into normal training data and $\bm{f}_{b}^{'}$ is the retrained recommender with recommender training loss function $\mathcal{L}_{\mathit{rec}}$ as:
        \begin{equation}
            \mathcal{\bm{Z}}^* = \arg \max_{\substack{\mathcal{\bm{Z}}}} \sum_{i = 1}^{|\mathcal{\bm{X}|}} \bm{E}_{t}(\bm{f}_{b}^{'}(\bm{x}^{(i)})) \quad \text{s.t.} \quad \bm{f}_{b}^{'} = \arg \min_{\substack{\bm{f}_{b}}} \mathcal{L}_{\mathit{rec}}(\bm{f}_{b}, \mathcal{\bm{Z}}\cup \mathcal{\bm{X}}).
        \end{equation}
        
        % \begin{equation}
        %      \min_{\substack{\bm{f}_{b}}} \Big \{ \mathcal{L}_{\mathit{rec}}(\bm{f}_{b}, [\bm{X}; \bm{Z}]) + \max_{\substack{\bm{Z}}} \{ \sum_{i = 1}^{|\bm{X}|} \bm{E}_{t}(\bm{f}_{b}( [\bm{X}_i; \bm{Z}_i]) \} \Big\}
        % \end{equation}
\end{itemize}

\section{Methodology}
\label{sec:method}

%In this section, approaches for our experiments are introduced. 
% We include details of 
%the framework implementation on sequential recommmenders as our framework in
% two parts: model extraction (\Cref{sec:extraction}) and downstream attacks (\Cref{sec:attack}). Different algorithms for each are introduced as follows.

% \label{sec:generation}

\subsection{Data-Free Model Extraction}

\label{sec:extraction}

To extract a black-box recommender 
%with
in a
data-free setting, we complete the process in two steps: (1)~\emph{data generation}, which generates input sequences $\mathcal{\bm{X}}$ and output ranking lists $\mathcal{\hat{\bm{I}}}^k$; (2)~\emph{model distillation}, using ($\mathcal{\bm{X}}$, $\mathcal{\hat{\bm{I}}}^k$) to minimize the difference between the black- and white-box recommender.

\subsubsection{Data Generation.}
\label{sec:generation}
Considering that we don't have access to the original training data and item statistics, a trivial solution is to make use of random data and acquire the model recommendations for later stages.
%Nevertheless, 
% However,
% random data cannot simulate real user behavior and might be characterized as a fake / malicious profile by detection algorithms, therefore, based on the features of sequential recommenders, we propose the following data generation methods for comparison:

\begin{itemize}
    \item \emph{Random:} Items are uniformly sampled from the input space to form sequences $\mathcal{\bm{X}}_\textit{rand}=\{{\bm{x}^{(i)}_{\textit{rand}}}\}_{i=1}^B$ where ${\bm{x}^{(i)}_{\textit{rand}}}$ is a generated sequence with identifier $i$ and $B$ is the budget size. Top-k items ($k=100$ in our experiments) are acquired from the victim recommender in each step to form the output result set, i.e.,~$\mathcal{\bm{\hat{I}}}^k_\textit{rand}=\{[\bm{f}_b({\bm{x}^{(i)}_{\textit{rand}}}_{[:1]}), \dots,\bm{f}_b({\bm{x}^{(i)}_{\textit{rand}}}_{[:T_i]})]\}^B_{i=1}$, where the $[:t]$ operation truncates the first $t$ items in the sequence 
    %(it is realistic that we can get a recommended 
    (corresponding to a recommendation
    list after each click). $T_i$ is the length of ${\bm{x}^{(i)}_\textit{rand}}$, where $T_i$ can be sampled from a pre-defined distribution or simply set as a fixed value. Following this strategy, we generate inputs and labels $(\mathcal{\bm{X_\textit{rand}}}, \mathcal{\hat{\bm{I}}}^k_\textit{rand})$ for model distillation
    % Labels of the random data have to be retrieved separately.
\end{itemize}

However, random data cannot simulate real user behavior sequences $\bm{x}_\textit{real}$ where sequential dependecies exist among different steps. To tackle this problem, we propose an autoregressive generation strategy. Inspired by autoregressive language models where generated sentences are similar to a `real' data distribution, and the finding that sequential recommenders are often trained in an autoregressive way~\cite{hidasi2015session, li2017neural, kang2018self}, we generate fake sequences autoregressively.

\begin{itemize}
    \item \emph{Autoregressive:}  To generate one sequence ${\bm{x}^{(i)}_\textit{auto}}$, a random item is sampled as the 
    %sequence starts 
    start item ${{\bm{x}^{(i)}_\textit{auto}}_{[1]}}$ (the $[t]$ operation selects the $t$-th item in sequence $\bm{x}$)
    and fed to a sequential recommender to get a 
    %recommended 
    recommendation
    list $\bm{f}_b({{\bm{x}^{(i)}_\textit{auto}}_{[:1]}})$. We repeat this step autoregressively i.e.,~${\bm{x}^{(i)}_\textit{auto}}_{[t]}=\texttt{sampler}(\bm{f}_b({{\bm{x}^{(i)}_\textit{auto}}_{[:t-1]}}))$ to generate sequences 
    %till 
    up to
    the maximum length $T_i$. Here $\texttt{sampler}$ is a method to sample one item from the given top-k list. In our experiment, sampling from the top-k items with monotonically decreasing probability performs similarly to uniform sampling, so we
    % simply 
    %choose uniform 
    favor
    sampling when selecting the next item from the top-k list. Accordingly, we generate $B$ sequences  $\mathcal{\bm{X}_\textit{auto}}=\{\bm{x}^{(i)}_\textit{auto}\}^B_{i=1}$ and record top-k lists, forming a dataset $(\mathcal{\bm{X}_\textit{auto}}, \hat{\mathcal{\bm{I}}}^k_\textit{auto})$ for 
    % further 
    model distillation.
\end{itemize}

\begin{figure*}[t]
  \centering
  \begin{subfigure}{0.43\textwidth}
    \includegraphics[width=1\linewidth]{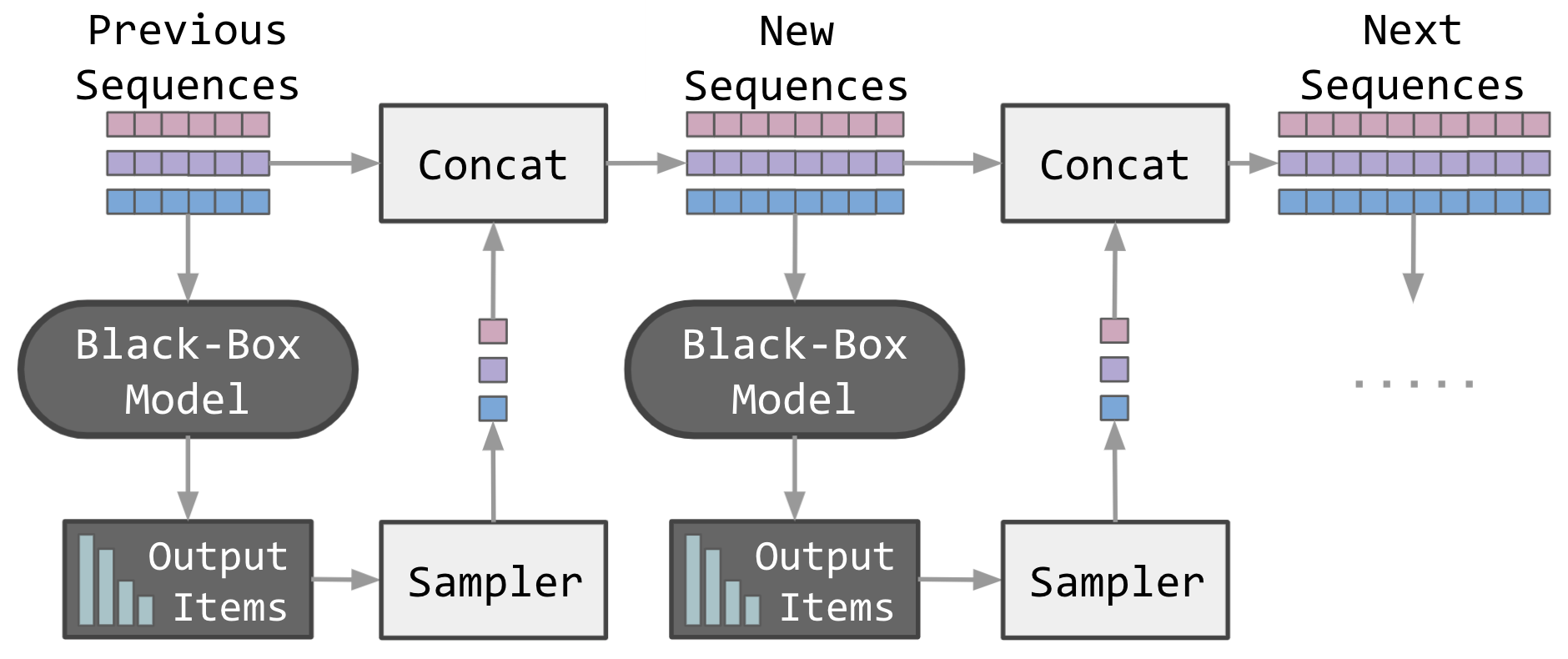}
    \caption{Autoregressive Data Generation}
    \label{fig:autoregressive}
  \end{subfigure}
  \hfill
  \begin{subfigure}{0.5\textwidth}
    \includegraphics[width=1\linewidth]{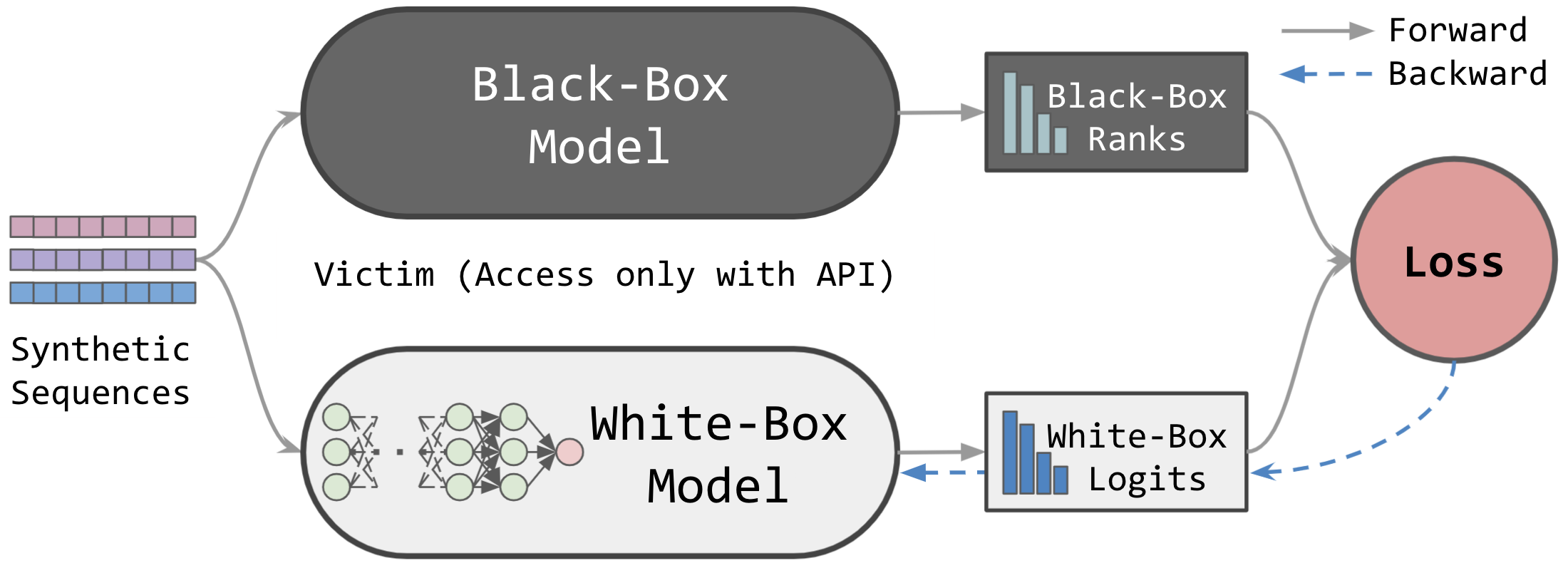}
    \caption{Model Extraction via Distillation}
    \label{fig:distillation}
  \end{subfigure}
  \hfill
  \vspace{-6pt}
  \caption{Autoregressive sequences and the use of synthetic data for model extraction.}
  \vspace{-16pt}
\end{figure*}

% \begin{figure*}[t]
%  \centering
%  \begin{subfigure}{0.4\textwidth}
%     \includegraphics[height=3cm]{paper/autoregressive.png}
%     \caption{Autoregressive Data Generation}
%     \label{fig:autoregressive}
%  \end{subfigure}
% %  \hfill
%  \begin{subfigure}{0.4\textwidth}
%     \includegraphics[height=3cm]{paper/distillation.png}
%     \caption{Model Extraction via Distillation}
%     \label{fig:distillation}
%  \end{subfigure}
%  \hfill
% \caption{Autoregressive sequences and the use of synthetic data for model extraction.}\vspace{-15pt}
% \end{figure*}

For autoregressive sequence generation, \Cref{fig:autoregressive} visually represents the process of accumulating data 
%length 
by repeatedly feeding current data to the recommender and appending current sequences with the sampled items from the model output. 
%\emph{Autoregressive} 
\emph{Autoregressive} method
is beneficial as: (1) Generated data is more diverse and more representative of real user behavior. 
%Especially, 
In particular,
sampling instead of choosing the first recommended item will help to build more diversified and `user-like' distillation data; (2) 
Since API queries are limited, 
%\emph{autoregressive} 
\emph{autoregressive} method
generates data by resembling real data distribution can obtain higher-quality data to train a surrogate model effectively; (3) 
%\emph{Autoregressive} 
%\emph{Autoregression}
It
%performs like numbers of `real' users wander on platforms by harmlessly `interacting with recommended items', 
resembles the behavior of real users
so 
%it 
is hard to detect.
Nevertheless, 
\emph{autoregressive} method
does not exploit output rankings and item properties like similarity
% and popularity 
due to restricted data access, which could limit the full utilization of a sequential recommender.

% Nevertheless, autoregressive data starts from randomly sampled elements and doesn't consider the influence of item popularity, which could limit the full utilization of a sequential recommender.

\subsubsection{Model Distillation.}
We use model distillation~\cite{hinton2015distilling} to minimize the difference between $\bm{f}_w$ and $\bm{f}_b$ by training with generated data $(\mathcal{\bm{X}}, \hat{\mathcal{\bm{I}}}^k)$ (see~\Cref{fig:distillation}). To get the most out of a single sequence during distillation, we generate sub-sequences and labels to enrich training data; for an input sequence $\bm{x}=[x_1, x_2, \ldots, x_T]$ from $\mathcal{\bm{X}}$, we split it into $T - 1$ entries of $\bm{x}_{[:2]}, \bm{x}_{[:3]}, \dots, \bm{x}_{[:T]}$ following training strategies in \cite{hidasi2015session, li2017neural}. 

% The next-item recommendations of subsequences can be retrieved and saved in each step during autoregressive generation, while black-box recommendations of random sequences are acquired from the black-box model separately after its sampling process.

% Their respective labels 
% %could 
% can
% be retrieved and saved during autoregressive data generation, while labels of random sequences would have to be acquired separately from the black-box model.

% Given the generated data, a white-box model is trained with distillation, this helps transfer the knowledge of the black-box model to an equally- or smaller-sized white-box model which could be accessed by adversaries. 

Compared to traditional model distillation~\cite{hinton2015distilling, Krishna2020Thieves}, where a model is distilled from 
%label 
%predicted 
predicted label
\emph{probabilities}, 
%our uniqueness is we 
in our setting we
only have top-k ranking list instead of 
%full 
probability distributions. So we propose a method to distill the model with a ranking loss. We can access the white-box model $\bm{f}_w$ output scores to items from the black-box top-k list $\hat{\bm{I}}=\bm{f}_b(\bm{x})$, which is defined as $\hat{\bm{S}}_w^k = [\bm{f}_{w}(\bm{x})_{[\hat{\bm{I}}_{[i]}^k]}]^k_{i=1}$
% $\hat{\bm{S}}_w^k = [\bm{f}_{w}(\bm{x})_{[\hat{\bm{I}}_{b[i]}^k]}]^k_{i=1}$
. For example, for $\hat{\bm{I}}^k = [25, 3, \ldots, 99]$
% $\hat{\bm{I}}_b^k = [25, 3, \ldots, 99]$
, $\hat{\bm{S}}_w^k = [\bm{f}_{w}(\bm{x})_{[25]}, \bm{f}_{w}(\bm{x})_{[3]}, \ldots, \bm{f}_{w}(\bm{x})_{[99]}]$). We also sample $k$ negative items uniformly and retrieve their scores as $\hat{\bm{S}}_{\mathit{neg}}^k$. We design a pair-wise ranking loss $\mathcal{L}_{\mathit{dis}}$ to measure the distance between black-box and white-box outputs:
\begin{equation}
    \mathcal{L}_{\mathit{dis}} = \frac{1}{k-1} \sum_{i=1}^{k-1} \max(0, \hat{\bm{S}}_{w [i+1]}^k - \hat{\bm{S}}_{w [i]}^k + \lambda_1) + \frac{1}{k} \sum_{i=1}^{k} \max(0, \hat{\bm{S}}_{neg [i]}^k - \hat{\bm{S}}_{w [i]}^k + \lambda_2).
\end{equation}

The loss function consists of two terms. The first term emphasizes ranking by computing a marginal ranking loss between all neighboring item pairs in $\hat{\bm{S}}_w^k$, which maximizes the probability of ranking positive items in 
%identical 
the same
order as the black-box recommender system. The second term punishes negative samples when they have higher scores than the top-k items, such that the distilled model learns to `recall' similar top-k groups for next-item recommendation. $\lambda_1$ and $\lambda_2$ are two margin values to be empirically set as hyperparameters. % \add{Here we assume can be approximated by a uniform distribution}

% Inspired by ranking losses and pair-wise training \cite{rendle2012bpr, tang2018ranking}, we design an empirical pair-wise ranking loss $\mathcal{L}_{\mathit{dis}}$ to measure the distance between black-box and white-box outputs, the loss function consists of two terms, the first term computes a marginal ranking loss between all neighboring item pairs in $\hat{\bm{S}}_w^k$, while the second term punishes negative samples if they have higher scores than top-k recommendations, $\lambda_1$ and $\lambda_2$ are two margin values to be empirically set as hyperparameters:

\subsection{Downstream Attacks}

\label{sec:attack}

%In this subsection,
To investigate whether attacks can be transfered
%ies 
from the trained white-box model $\bm{f}^*_w$\footnote{
%Without ambiguity, 
%To avoid ambiguity,
We directly use $\bm{f}_w$ to represent the trained white-box $\bm{f}^*_w$ below.} to the black-box model $\bm{f}_b$, we introduce two model attacks against sequential recommender systems: \emph{profile pollution} attacks and \emph{data poisoning} attacks, see~\Cref{fig:intro} for illustration of the two attack scenarios.

% The former allows the attacker to inject fake interactions after the original sequence, while the latter aims to generate certain amounts of malicious profiles in order to bias the system's output after retraining. 

\begin{figure*}[t]
  \centering
  \begin{subfigure}{0.48\textwidth}
  \centering
    \includegraphics[width=0.9\linewidth]{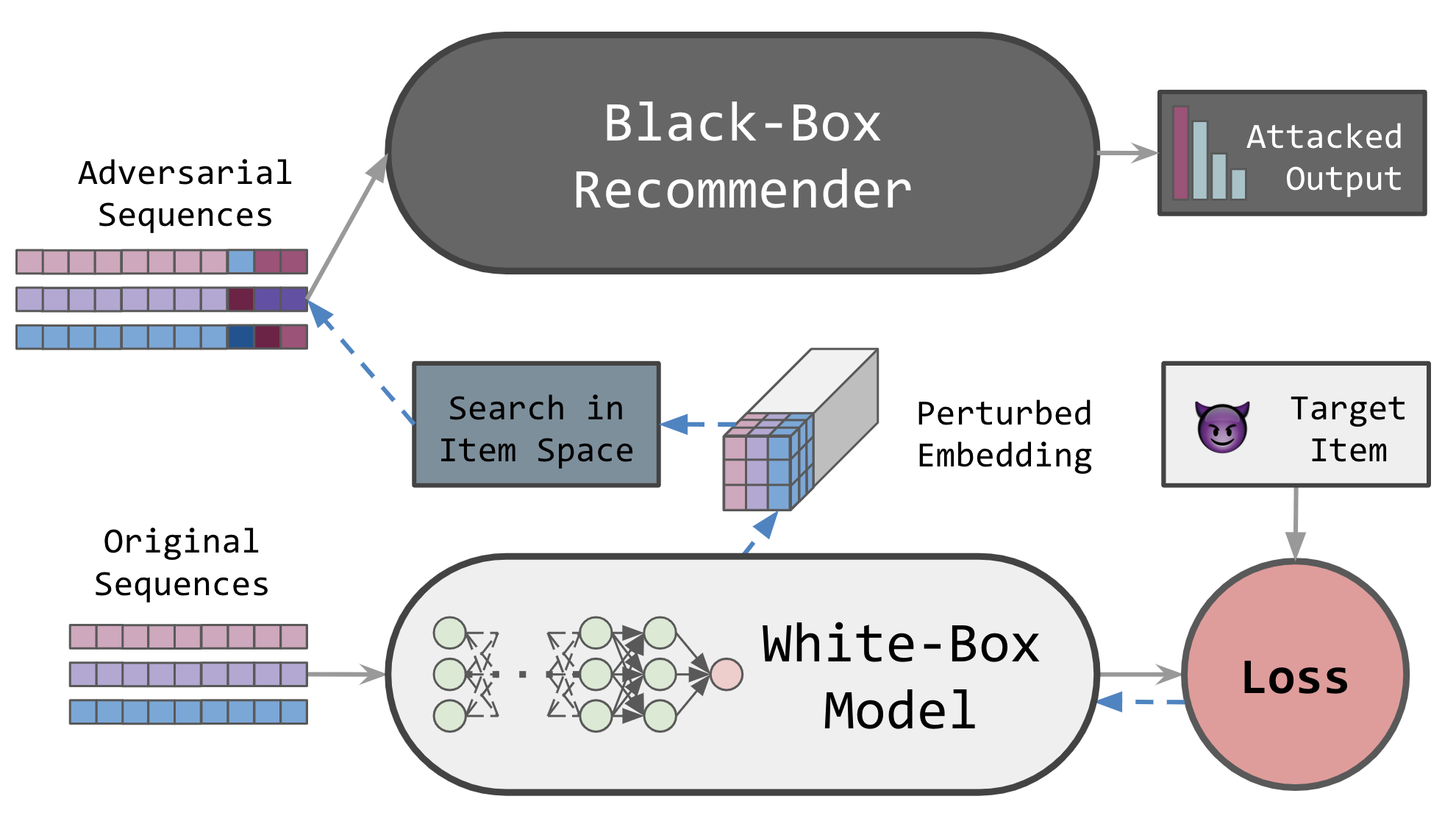}
    \caption{Profile pollution attack with white-box model}
    \label{fig:attacking}
  \end{subfigure}
  \hfill
  \begin{subfigure}{0.51\textwidth}
  \centering
    \includegraphics[width=0.9\linewidth]{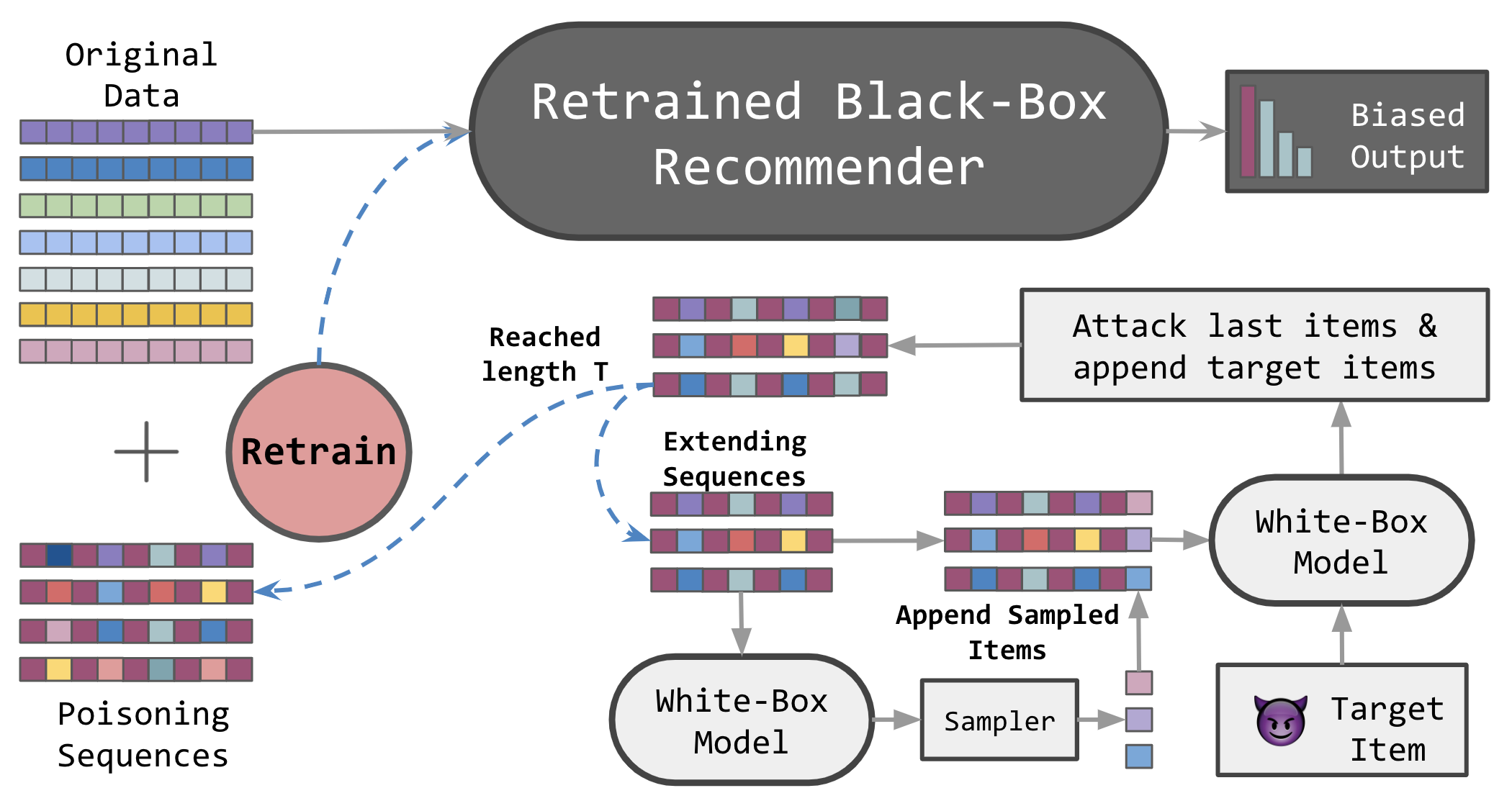}
    \caption{Data poisoning attack via adversarial co-visitation}
    \label{fig:poison}
  \end{subfigure}
  \hfill
  \vspace{-6pt}
  \caption{Two scenarios: profile pollution attack and data poisoning attacks.}
  \vspace{-16pt}
\end{figure*}

\subsubsection{Profile Pollution Attack}
As described in~\Cref{fig:attacking}, we perform profile pollution attacks to promote item exposure and use%
%the proposed
~\Cref{alg:attack} to construct the manipulated sequence in the input space. 
% Similar to previous works~\cite{zeller2008cross, xing2013take}, we assume the victim profile is accessible to adversaries, and adversarial items could be injected after the original sequence in order to maximize 
%the
Because we can access the gradients in white box models, %it enables us 
we are able
to append `adversarial' items by extending adversarial example techniques (e.g.~Targeted Fast Gradient Sign Method (T-FGSM)~\cite{goodfellow2014explaining}) from continuous feature space (e.g.~image pixel values) to discrete item space (e.g.~item IDs), with an assumption that the optimal item is `close' to target item in embedding space.
Therefore we can achieve user-specific item promotion without the black-box recommender being retrained as below.

\xhdr{Step 1: Compute Gradients at the Embedding Level.}
Given the user history $\bm{x}$, we construct the corrupted sequence $\bm{z}$ by appending adversarial items after $\bm{x}$. We first initialize $\bm{z}$
to be the
same as $\bm{x}$ and append target items to it. With $\bm{f}_{w, m}$ from the previous step, we feed the embedded input $\tilde{\bm{z}}$ to the model and compute backward gradients w.r.t.~the input embeddings using a cross-entropy loss, where the target item {$t$} is used as a label: $\nabla = \nabla_{\tilde{\bm{z}}} \mathcal{L}_{\mathit{ce}} (\bm{f}_{w, m}(\tilde{\bm{z}}), t)$. 

% We first initialize $\bm{z}$ with the original user sequence concatenated with target items appended to it. 

% The backward gradients on the embedding Level is therefore:

% \begin{equation}
%     \nabla = \nabla_{\tilde{\bm{x}}} \mathcal{L}_{\mathit{ce}} (\bm{f}_{w, m}(\tilde{\bm{x}}), y).
% \end{equation}

% The sign of the gradients will be multiplied with a factor $\epsilon$ and then added or subtracted from the original embedding, depending on whether the attack is untargeted or targeted with $y$ representing the label or target item:
% \begin{itemize}
%     \item FGSM: 
%     \begin{equation}
%         \tilde{\bm{x}}^{adv} = \tilde{\bm{x}} + \tilde{\bm{\delta}} = \tilde{\bm{x}} + \epsilon \mathit{sign}(\nabla \mathcal{L}(\bm{F_E}(\tilde{\bm{x}}), y))
%     \end{equation}
    
%     \item T-FGSM: 
%     \begin{equation}
%         \tilde{\bm{x}}^{adv} = \tilde{\bm{x}} + \tilde{\bm{\delta}} = \tilde{\bm{x}} - \epsilon \mathit{sign}(\nabla \mathcal{L}(\bm{F_E}(\tilde{\bm{x}}), y))
%     \end{equation}
% \end{itemize}

\begin{algorithm}[t]
\caption{Adversarial Item Search for Profile Pollution}
\label{alg:attack}
\SetAlgoLined
\textbf{Input}
sequence $\bm{x}$, target $t$, expected length of polluted sequence $T$, white-box model $\bm{f}_{w}$ (i.e.~$\bm{f}_{w, e}$, $\bm{f}_{w, m}$), $\epsilon$ and $n$\;
\textbf{Output}
polluted sequence $\bm{z}$\;
initialize $\bm{z}$ with $\bm{z} = \bm{x}$\;
\While{length of $\bm{z} < T$}{
  \text extend $\bm{z}$ by appending target item:
  $\bm{z}  \leftarrow [\bm{z}; t]$\;
  transform sequence into sequence embeddings: $\tilde{\bm{z}} = \bm{f}_{w, e}(\bm{z})$\;
  compute backward gradients using cross entropy $\nabla = \nabla_{\tilde{\bm{z}}} \mathcal{L}_{\mathit{ce}} (\bm{f}_{w, m}(\tilde{\bm{z}}), t)$\;
  compute cosine similarity scores $\bm{S}$ between $\tilde{\bm{z}} - \epsilon \textit{sign}(\nabla)$ and $\bm{f}_{w, e}(i)\ \forall i \in \mathcal{I}$\;
  select $n$ candidates $\bm{C}$ with highest $\bm{S}$ scores, exclude $t$ if repeated\;
  replace $t$ in $\bm{z}$ with $c \in \bm{C}$ and keep $\bm{z}$ with highest $\bm{f}_{w}(\bm{z})_{[t]}$\;
}

\end{algorithm}

\xhdr{Step 2: Search for Adversarial Candidates.}
Based on $\bm{z}$ and $\nabla$ from the previous step, we first perform T-FGSM to compute the perturbed embeddings $\tilde{\bm{z}}' = \tilde{\bm{z}} - \epsilon \textit{sign}(\nabla)$, then the cosine similarity of the embedded injection item in $\tilde{\bm{z}}'$ is computed with all item embeddings across $\mathcal{I}$. We select $n$ adversarial candidates with the highest cosine similarity. These items are tested with $\bm{f}_w$ such that the candidate leading to the highest ranking score of target
% resulting in highest target exposure 
is kept as the adversarial item. It can be repeated for multiple injection items for better attack performance, and to avoid disproportionate target items in injection, we require target items not appear continuously.

\subsubsection{Data Poisoning}

\begin{algorithm}[t]
\caption{Adversarial Profile Generation for Data Poisoning}
\label{alg:poison}
\SetAlgoLined
\textbf{Input}
target $t$, expected length $T$, white-box model $\bm{f}_{w}$ (i.e.~$\bm{f}_{w, e}$, $\bm{f}_{w, m}$), $\epsilon$ and $n$\;
\textbf{Output}
adversarial profile $\bm{z}$\;
\text initialize $\bm{z}$ with $\bm{z} = [t]$\;
\While{length of $\bm{z} < T$}{
  sample item $z_{\mathit{temp}}$ in $\mathcal{I}$ and append to $\bm{z}$: $\bm{z} = [\bm{z}; z_{\mathit{temp}}]$\;
  transform sequence into sequence embeddings: $\tilde{\bm{z}} = \bm{f}_{w, e}(\bm{z})$\;
  compute backward gradients using cross entropy $\nabla = \nabla_{\tilde{\bm{z}}} \mathcal{L}_{\mathit{ce}} (\bm{f}_{w, m}(\tilde{\bm{z}}), t)$\;
  compute cosine similarity scores $\bm{S}$ between $\tilde{\bm{z}} - \epsilon \textit{sign}(\nabla)$ and $\bm{f}_{w, e}(i)\ \forall i \in \mathcal{I}$\;
  select $n$ candidates $\bm{C}$ with lowest $\bm{S}$ scores, exclude $t$ if repeated\;
  sample item in $\bm{C}$ and replace $z_{\mathit{temp}}$ in $\bm{z}$\;
  append target item $\bm{z} \leftarrow [\bm{z}; t]$\;
}
\end{algorithm}

Data poisoning attacks 
%aim to achieve such attacks with 
operate via
fake profile injection, to promote target item exposure as much as possible (after retraining with the fake 
%profiles 
and normal profiles). We propose a simple adversarial strategy (visualized in~\Cref{fig:poison}) to generate poisoning data with white-box model $\bm{f}_w$. The intuition behind our poisoning data generation is that 
%even the 
%\emph{most irrelevant items sequence} should be labeled as \emph{the next item is the target}.
\emph{the next item should be the target even given sequences of seemingly irrelevant items.}

% and use alternating items pairs (e.g.~[item 1, target, item 3, target, ...]) to avoid disproportionate target items.

% avoid the trivial solution of repeating target items. 

% We consider different amounts of injection sequences and evaluate the respective retrained model. 

In this case, we follow co-visitation approach~\cite{yang2017fake, song2020poisonrec} and apply adversarial example techniques~\cite{goodfellow2014explaining} to generate poisoning data. (1)~We consider one poisoning sequence using alternating items pairs (e.g.~$\bm{z}=$[target, ${z}_2$, target, ${z}_4$, target, ...]); (2)~We try to find \emph{irrelevant/unlikely items} to fill ${z}_{2d}$ in $\bm{z}$. In detail, we use a similar approach to~\Cref{alg:attack}, where the generation process first computes backward gradients similarly 
%with 
to the
cross entropy loss and T-FGSM. However, in narrowing candidate items we choose from $\mathcal{I}$
%the 
% popular items $\mathcal{P}$
with the lowest similarity scores to $\tilde{\bm{z}} - \epsilon \textit{sign}(\nabla)$ (instead of the highest); (3)~We repeat (1) and (2) to generate the poisoning data; details 
%of 
can be found in~\Cref{alg:poison}. 

Note that though the alternating pattern of co-visitation seems detectable, we can control the proportion of target items (apply noise or add `user-like' generated data) to avoid this rigid pattern and make it less noticeable. Gradient information from the white-box model also empowers more tailored sequential-recommendation poisoning methods.

\section{Experiments}

\subsection{Setup}
\subsubsection{Dataset} We use three popular recommendation datasets (see~\Cref{tab:datasets}) to evaluate our methods:
%, which are 
Movielens-1M (ML-1M)~\cite{harper2015movielens}, Steam~\cite{mcauley2015image} and Amazon Beauty~\cite{ni-etal-2019-justifying}. We follow the preprocessing in BERT4Rec~\cite{sun2019bert4rec} to process the rating data into implicit feedback. We follow SASRec~\cite{kang2018self} and BERT4Rec~\cite{sun2019bert4rec} to hold out the last two items in each sequence 
%as 
for
validation and testing,
%item, 
and the rest for training.
%set. 
We set black-box API returns to be top-100 recommended items.

\begin{table}[t]
\parbox{.45\linewidth}{
\small
\centering
% \begin{tabular}{cccccc}
% \toprule
% \textbf{Datasets} & \textbf{\#users} & \textbf{\#items} & \textbf{\#actions} & \textbf{Avg. Len} & \textbf{Sparsity} \\ \midrule
% \textbf{ML-1M}    & 6,040            & 3,416            & 1.0m               & 163.5                & 95.15\%           \\
% \textbf{Steam}     & 334,730         & 13,047           & 3.7m               & 11.1                  & 99.92\%           \\ 
% \textbf{Beauty}   & 40,226           & 54,542           & 0.35m              & 7.6                  & 99.98\%           \\
% \bottomrule
% \end{tabular}
\begin{tabular}{cccccc}
\toprule
\textbf{Datasets} & \textbf{Users} & \textbf{Items} & \textbf{Avg. len} & \textbf{Max. len} & \textbf{Sparsity} \\ \midrule
\textbf{ML-1M}    & 6,040          & 3,416          & 166               & 2277              & 95.16\%           \\
\textbf{Steam}    & 334,542        & 13,046         & 13                & 2045              & 99.90\%           \\
\textbf{Beauty}   & 40,226         & 54,542         & 9                 & 293               & 99.98\%           \\ \bottomrule
\end{tabular}
\caption{Data Statistics} \vspace{-15pt}
\label{tab:datasets}
}
\hfill
\parbox{.45\linewidth}{
\small
\centering
\begin{tabular}{ccc}
\toprule
\textbf{Model}    & \textbf{Basic Block} & \textbf{Training Schema} \\ \midrule
\textbf{NARM~\cite{li2017neural}}     & GRU            & Autoregressive    \\
\textbf{SASRec~\cite{kang2018self}}   & TRM    & Autoregressive    \\
\textbf{BERT4Rec~\cite{sun2019bert4rec}} & TRM    & Auto-encoding     \\
\bottomrule
\end{tabular}
\caption{Sequential Model Architecture}\vspace{-16pt}
\label{tab:model}
}
\end{table}

\subsubsection{Model.} To evaluate the performance of our attack, we implement a model extraction attack on three representative sequential recommenders, including our PyTorch implementations of NARM~\cite{li2017neural}, BERT4Rec~\cite{sun2019bert4rec} and SASRec~\cite{kang2018self}, with different basic blocks and training schemata, shown in \Cref{tab:model}. 

\begin{itemize}
    \item \emph{NARM} is an attentive recommender, containing an embedding layer, gated recurrent unit (GRU)~\cite{cho2014learning} as a global and local encoder, an attention module to compute session features and a similarity layer, which outputs the most similar items to the session features as recommendations~\cite{li2017neural}.
    \item \emph{SASRec} consists of an embedding layer that includes both the item embedding and the positional embedding of an input sequence as well as a stack of one-directional transformer (TRM) layers, where each transformer layer contains a multi-head self-attention module and a position-wise feed-forward network~\cite{kang2018self}. 
    \item \emph{BERT4Rec} has an architecture similar to \emph{SASRec}, but using a bidirectional transformer and an
    %the 
    auto-encoding (masked language modeling)
    %masking 
    task for training~\cite{sun2019bert4rec}.
\end{itemize}

\subsubsection{Implementation Details.} Given a user sequence $\bm{x}$ with length $T$, we follow~\cite{kang2018self, sun2019bert4rec} to use $\bm{x}_{[:T-2]}$  as training data and use the last two items for validation and testing respectively. We use hyper-parameters from grid-search and suggestions from original papers~\cite{li2017neural, kang2018self, jin2019bert}. For reproducibility, we summarize important training configurations in~\Cref{tab:config}. Additionally, all models are trained using Adam~\cite{kingma2014adam} optimizer with weight decay 0.01, learning rate 0.001, batch size 128 and 100 linear warmup steps. We follow~\cite{kang2018self, sun2019bert4rec} to set allowed sequence lengths of ML-1M, Steam and Beauty as $\{200, 50, 50\}$ respectively, which are also applied as our generated sequence lengths. We follow~\cite{tang2020revisiting} to use 1\% of a real profile's size as a poisoning profile size. Code and data are released\footnote{\href{https://github.com/Yueeeeeeee/RecSys-Extraction-Attack}{https://github.com/Yueeeeeeee/RecSys-Extraction-Attack}}.

\begin{table}[h]
\renewcommand\arraystretch{1.15} 
\footnotesize

\begin{tabular}{ccccc}
\toprule
\textbf{Phase}                      & \textbf{Model} & \textbf{Config. on \{ML-1M, Steam, Beauty\}}                    & \textbf{Phase}   & \textbf{Config. on \{ML-1M, Steam, Beauty\}}    \\ \midrule
\multirow{3}{*}{\textbf{\shortstack{Black-box \\Training}}} & N              & GRU ly=1; dr=\{0.1, 0.2, 0.5\}                               & \textbf{Model Extraction} & $\lambda_1,\lambda_2$: \{(0.75, 1.5), (0.5, 1.0), (0.5, 0.5)\} \\
                                    & S              & TRM ly=2; h=2; dr=\{0.1, 0.2, 0.5\}                       & \textbf{Profile Pollution} & Append items: \{10, 2, 2\}, $\epsilon$ = 1.0, n = 10                      \\
                                    & B              & TRM ly=2; h=2; dr=\{0.1, 0.2, 0.5\}; mp=\{0.2, 0.2, 0.6\} & \textbf{Data Poisoning}   & Injected users: \{60, 3345, 402\}, $\epsilon$ = 1.0, n = 10               \\ \bottomrule          
\end{tabular}

\caption{Configurations. N:NARM, S:SASRec, B:Bert4Rec, ly:layer, h:attention head, dr:dropout rate, mp:masking probability.}\vspace{-16pt}
\label{tab:config}
\end{table}

\iffalse
\begin{figure*}[h]
    \centering
    \includegraphics[width=1\linewidth]{arch_comparison.png}
    \caption{The architectures of NARM, SASRec and BERT4Rec}\label{fig:architecture}
\end{figure*}
\fi

\subsubsection{Evaluation Protocol}

We follow SASRec~\cite{kang2018self} to accelerate evaluation by uniformly sampling 100 negative items for each user. Then we rank them with the positive item and report the average performance on these 101 testing items. Our Evaluation focuses on two aspects:
\begin{itemize}
    \item \textbf{Ranking Performance}: We to use truncated \emph{Recall@K} that is equivalent to Hit 
    %Ratio 
    Rate
    (\emph{HR@K}) in our evaluation, and Normalized Discounted Cumulative Gain (\emph{NDCG@K}) to measure the ranking quality following SASRec~\cite{kang2018self} and BERT4Rec~\cite{sun2019bert4rec}, higher is better.
    \item \textbf{Agreement Measure}: We define \emph{Agreement@K} (\emph{Agr@K}) to evaluate the output similarity between the black-box model and our extracted white-box model:
    \begin{equation}
        \mathit{Agreement}@K = \frac{|\mathit{B}_\text{topK}\cap \mathit{W}_\text{topK}|}{K}, 
    \end{equation}
    where $\mathit{B}_\text{topK}$ is the top-K predicted list from the black-box model and $\mathit{W}_\text{topK}$ is from our white-box model. We report average \emph{Agr@K} with $K=1, 10$ to measure the output similarity.
\end{itemize}

\begin{table}
% []
\small
\begin{tabular}{lccccccccccc}
\toprule
                              &          & \multicolumn{2}{c}{Black-Box}   & \multicolumn{4}{c}{White-Box-Random}                     & \multicolumn{4}{c}{White-Box-Autoregressive}                      \\ 
                              \cmidrule(l){3-4} \cmidrule(l){5-8} \cmidrule(l){9-12} 
Dataset                       & Option   & N@10           & R@10           & N@10  & R@10           & Agr@1          & Agr@10         & N@10           & R@10           & Agr@1          & Agr@10         \\ \midrule
\multirow{3}{*}{ML-1M}        & NARM     & \textbf{0.625} & \textbf{0.820} & 0.598 & 0.809          & 0.389          & 0.605          & 0.615          & 0.812          & \textbf{0.571} & \textbf{0.747} \\
                              & SASRec   & \textbf{0.625} & \textbf{0.817} & 0.578 & 0.796          & 0.270          & 0.516          & 0.602          & 0.802          & \textbf{0.454} & \textbf{0.662} \\
                              & BERT4Rec & \textbf{0.602} & \textbf{0.806} & 0.565 & 0.794          & 0.241          & 0.488          & 0.571          & 0.791          & \textbf{0.339} & \textbf{0.593} \\ \midrule
\multirow{3}{*}{Steam}        & NARM     & \textbf{0.628} & 0.848          & 0.625 & \textbf{0.849} & 0.679          & 0.642          & 0.601          & 0.806          & \textbf{0.743} & \textbf{0.722} \\
                              & SASRec   & \textbf{0.627} & \textbf{0.850} & 0.579 & 0.802          & 0.434          & 0.556          & 0.593          & 0.805          & \textbf{0.668} & \textbf{0.702} \\
                              & BERT4Rec & \textbf{0.622} & \textbf{0.846} & 0.609 & 0.838          & 0.199          & 0.490          & 0.585          & 0.793          & \textbf{0.708} & \textbf{0.667} \\ \midrule
\multirow{3}{*}{{Beauty}} & NARM     & \textbf{0.356} & \textbf{0.518} & 0.319 & 0.477          & \textbf{0.356} & \textbf{0.511} & 0.272          & 0.380          & 0.344          & 0.425          \\
                              & SASRec   & 0.344          & 0.494          & 0.304 & 0.459          & 0.251          & 0.213          & \textbf{0.347} & \textbf{0.505} & \textbf{0.343} & \textbf{0.357} \\
                              & BERT4Rec & \textbf{0.349} & \textbf{0.515} & 0.200 & 0.352          & 0.026          & 0.043          & 0.300          & 0.454          & \textbf{0.178} & \textbf{0.291} \\ \bottomrule
\end{tabular}
\caption{Extraction performance under identical model architecture and 5k budget, with Black-box original performance.}
\label{tab:distill}
\vspace{-8pt}
\end{table}

\begin{figure}[t]
  \vspace{-10pt}
  \begin{subfigure}[t]{.231\columnwidth}
  % include first image
    \includegraphics[width=\linewidth]{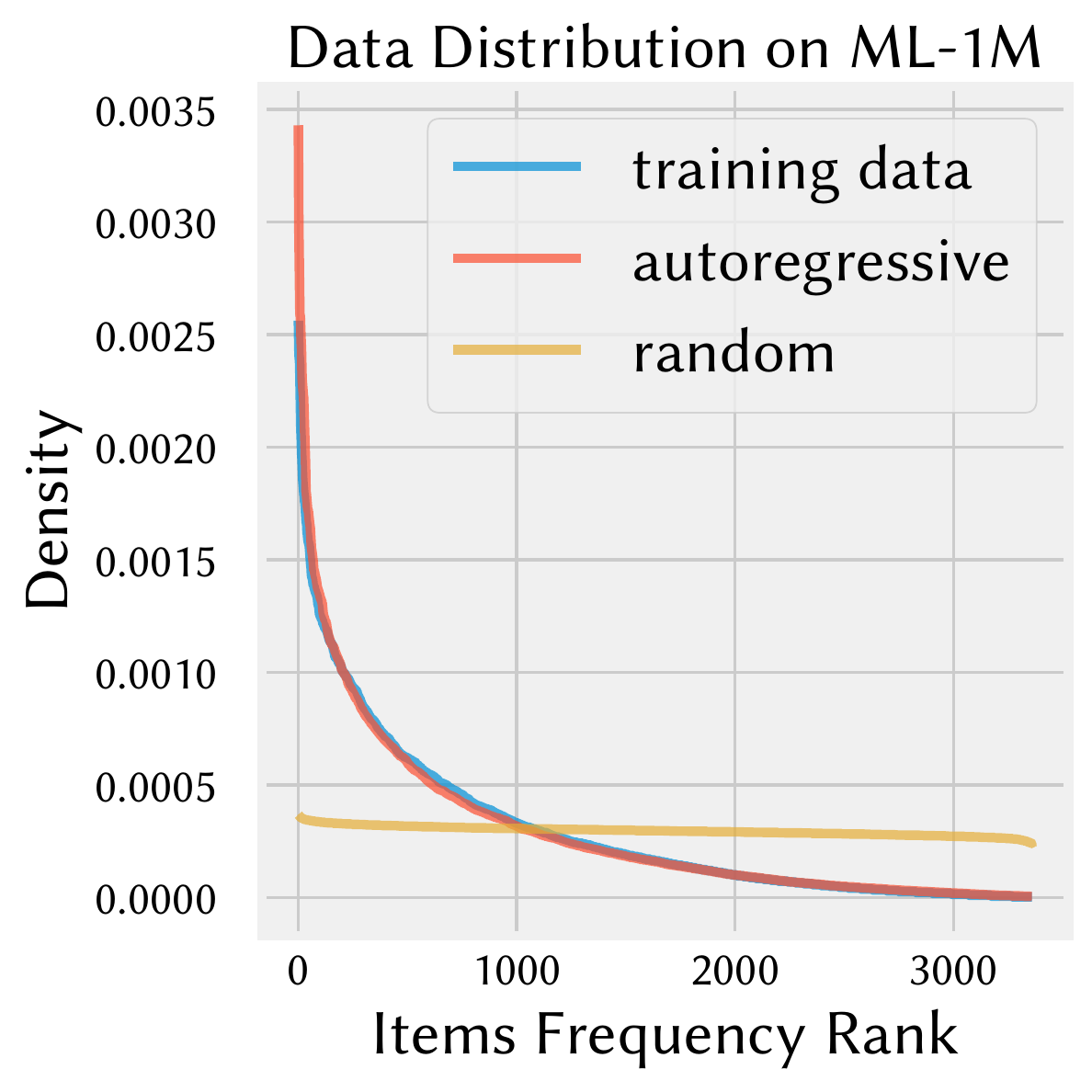}
  \caption{Data Distribution for original and generated data.}
  \label{fig:dist}
  \end{subfigure}
  \hspace{3mm}
  \begin{subfigure}[t]{.713\columnwidth}
  % include second image
    \includegraphics[width=0.3\linewidth]{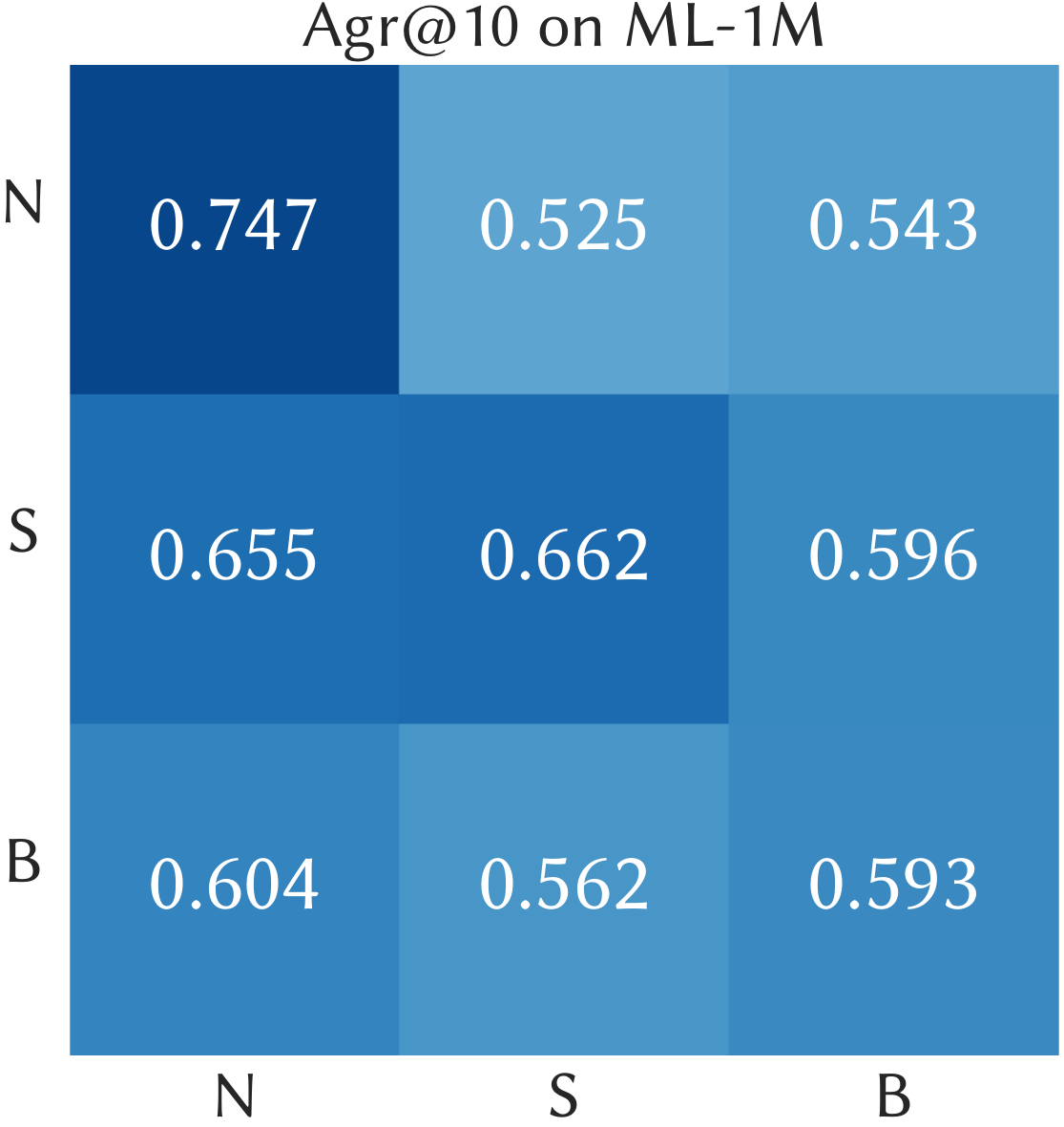}  
    \hspace{1mm}
    \includegraphics[width=0.3\linewidth]{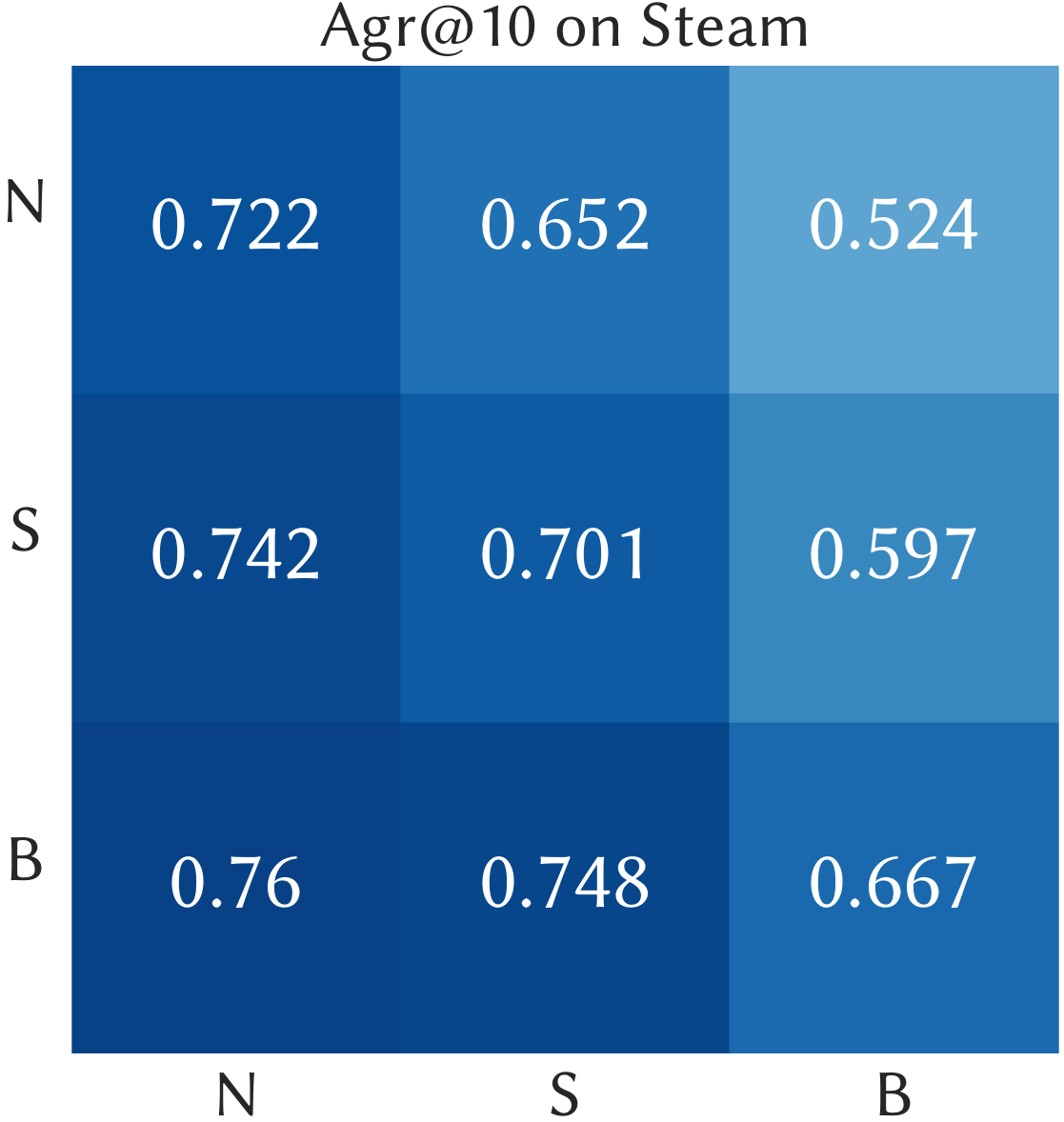}  
    \hspace{1mm}
    \includegraphics[width=0.3\linewidth]{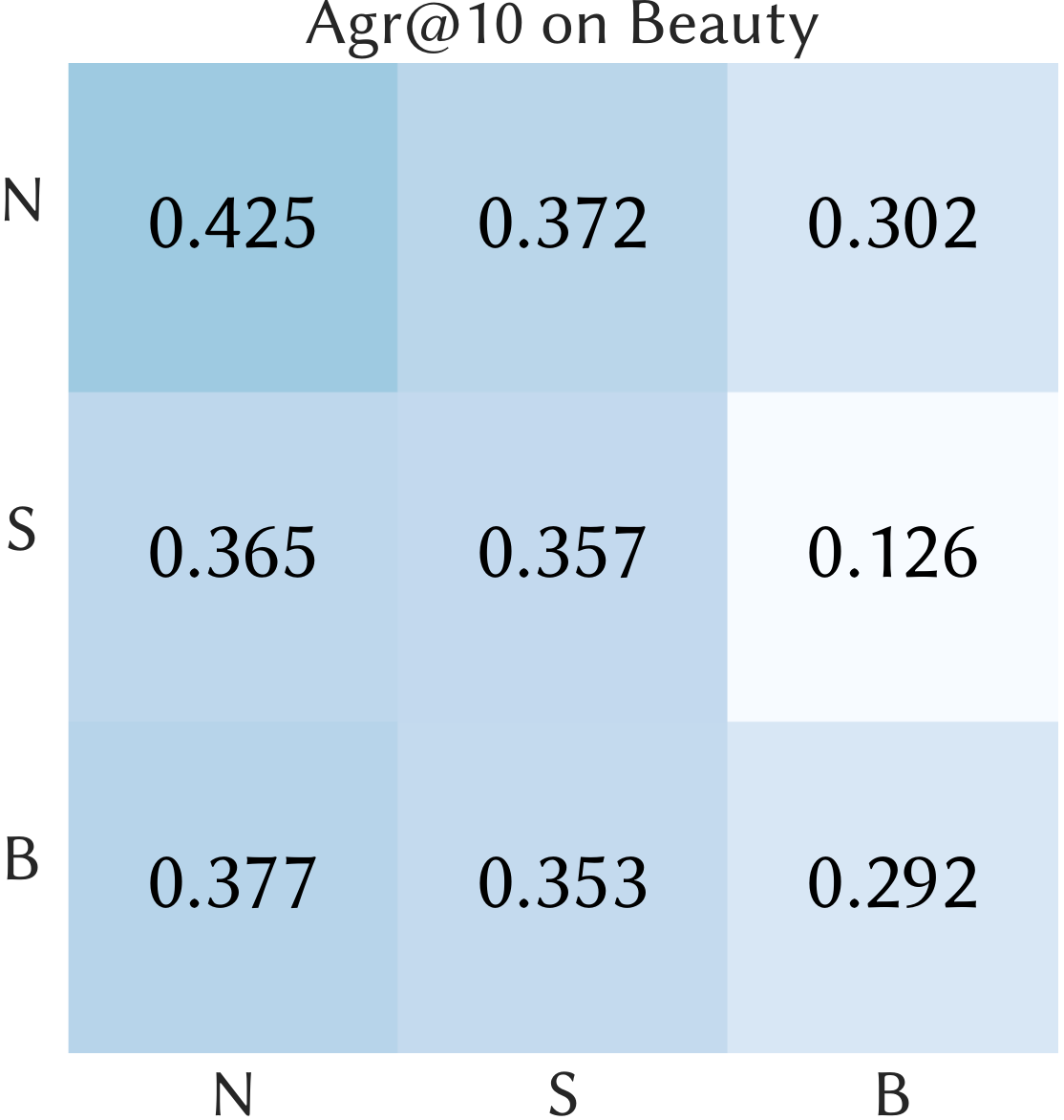}  
    \caption{Cross model extraction results.  Horizontal / vertical axes represent white-box / black-box model architectures. Darker colors represents larger values of Agr@10.}
    \label{fig:cross}
  \end{subfigure}
  \vspace{-6pt}
  \caption{Data distributions and cross model extraction results.}
  \vspace{-16pt}
\end{figure}

\subsection{RQ1: Can We Extract Model Weights without Real Data?}
\label{exp:rq1}

\subsubsection{Standard Model Extraction.} We evaluate two ways (\emph{random} and \emph{autoregressive} as mentioned in~\Cref{sec:generation}) of generating synthetic data with labels for model extraction. In standard model extraction, we assume the 
%knowledge of 
model architecture is known, so that the white-box model uses the 
%identical 
same
architecture as the black-box model (e.g.~SASRec$\rightarrow$SASRec) 
without real training data. We report results with a fixed budgets $B=5000$ in~\Cref{tab:distill}.

\xhdr{Observations.} From~\Cref{tab:distill} we have a few observations: 
(1)~Interestingly, 
%though 
without a
training set, \emph{random} and \emph{autoregressive} can achieve similar ranking 
%performances 
performance
(N@10 and R@10) as the \emph{black-box}. For example, compared with \emph{black-box} NARM on ML-1M, R@10 for \emph{random} drops 1.34\% and \emph{autoregressive} drops only 0.98\%. 
%Averagely, 
On average,
extracted recommenders' R@10 is about 94.54\% of the original. %  (0.686 + 0.683) / (2 * 0.724) 
Particularly, \emph{random} is trained on random data, but labels are retrieved from \emph{black-box} model, reflecting correct last-click relations. Last clicks help \emph{random} rank well, but agr@K is much poorer than \emph{autoregressive} (see~\Cref{tab:distill}).

(2)~\emph{Autoregressive} has significant advantages in narrowing the distance between the two recommenders in all datasets, with an average Agr@10 of $0.574$ against $0.452$ from randomly generated data. 
%In the case of NARM on Beauty, \emph{autoregressive} with budget $B=5000$ performs worse than $B=3000$ possibly due to overfitting synthetic user behavior, see \Cref{fig:budget}.
% The \emph{Recall} performance of the two distilled models in the test set is however highly similar, which is possibly related to the evaluation method based on the $100$ randomly sampled negative items. 
(3)~\Cref{fig:dist} shows \emph{autoregressive} resembles 
the
true training data distribution much better than \emph{random}, because \emph{autoregressive} generates data following 
%`interacting with recommended items', 
interactions with recommended items.
%which makes it look like real users and hard to be repressed. 
%Noting
%though
%that 
Although
%though 
sampling 
%with 
from the
popularity distribution can also resemble the original data distribution, it breaks the assumption that we have no knowledge about the training set, and 
%it 
cannot 
%resemble the 
capture
similarities 
%of 
from
sequential dependencies.
(4) We also note that the datasets have large differences in the distillation process. For example, relatively dense datasets with 
%plenty of 
many
user interactions like ML-1M and Steam increase the probability of correct recommendations. Extracted recommenders based on such data distributions sustain a high level of similarity with respect to the black-box output distribution, while in sparser data, it could lead to problems like higher divergence and worse recommendation agreement, which we will examine in the next subsection.
(5) Moreover, \Cref{tab:distill} indicates that NARM has the best overall capability of extracting black-box models, as NARM is able to recover most of the black-box models and both its similarity and recommendation metrics are among the highest. 
%On ML-1M and Steam datasets, NARM performs at least equally well or better in extraction experiments.
% , even for relatively sparse datasets like Beauty, NARM's performance maintains the same level, restoring at least $60\%$ and $80\%$ of the original \emph{R@10} accuracy under the $3000$ budget of random and autoregressive sequences, while SASRec and BERT4Rec experience different degrees of deterioration. 
As for SASRec and BERT4Rec, both architectures show satisfactory extraction results on 
the
ML-1M and Steam datasets, with SASRec showing slight improvements compared to BERT4Rec in most cases.
% , considering its overall higher similarity and better recommendation metrics.

\subsubsection{Cross Model Extraction.} Based on the analysis of different architectures, a natural question follows: which model would perform the best on a different black-box architecture? In this 
%sense, 
setting,
we adopt the same budget and conduct cross-extraction experiments to find out how a white-box recommender would differ in terms of similarity when distilling a different black-box architecture. 
%The following 
Cross-architecture model extraction is evaluated on these three datasets. 

\xhdr{Observations.} Results are visualized with heatmaps in \Cref{fig:cross}, where the horizontal / vertical axes represent
white-box / black-box architectures.
%architectures and vertical axis the black-box ones. 
The NARM model performs the best overall as a white-box architecture, successfully reproducing most target recommender systems with an average of $0.597$ agreement in the top-$10$ recommendations, compared to $0.548$ of SASRec and $0.471$ of BERT4Rec.

\subsection{RQ2: How do Dataset Sparsity and Budget Influence Model Extraction?}
\label{exp:rq2}

\begin{table}[t]
\parbox{.55\linewidth}{
\centering
\small
\begin{tabular}{lccccccc}
\toprule
                &       & \multicolumn{2}{c}{Black-Box} & \multicolumn{3}{c}{Ours} \\ \cmidrule(l){3-4} \cmidrule(l){5-7} 
{Model}            & {k-core} & {N@10}      & {R@10}     & {N@10}  & {R@10} & {Agr@10} \\ \midrule
\multirow{4}{*}{NARM}     & 5               & 0.360              & 0.536             & 0.331          & 0.488         & 0.559           \\
                          & 6               & 0.372              & 0.562             & 0.352          & 0.539         & 0.614           \\
                          & 7               & 0.386              & 0.630             & 0.369          & 0.597         & \textbf{0.726}  \\
                          & 8               & 0.347              & 0.597             & 0.362          & 0.643         & 0.690           \\ \midrule
\multirow{4}{*}{SASRec}   & 5               & 0.351              & 0.514             & 0.332          & 0.479         & 0.651           \\
                          & 6               & 0.380              & 0.558             & 0.373          & 0.547         & 0.744           \\
                          & 7               & 0.424              & 0.640             & 0.427          & 0.648         & 0.782           \\
                          & 8               & 0.415              & 0.675             & 0.410          & 0.672         & \textbf{0.791}  \\ \midrule
\multirow{4}{*}{BERT4Rec} & 5               & 0.346              & 0.509             & 0.351          & 0.520         & 0.561           \\
                          & 6               & 0.366              & 0.547             & 0.374          & 0.555         & 0.652           \\
                          & 7               & 0.402              & 0.643             & 0.399          & 0.650         & 0.682           \\
                          & 8               & 0.403              & 0.694             & 0.383          & 0.659         & \textbf{0.717}  \\ \bottomrule
\end{tabular}
\captionof{table}{Influence of data sparsity. Model extraction on $k$-core Beauty}
\label{tab:distillation_sparsity}
}
\hfill
\parbox{.43\linewidth}{
    \begin{subfigure}[b]{0.41\textwidth}
\includegraphics[width=\linewidth]{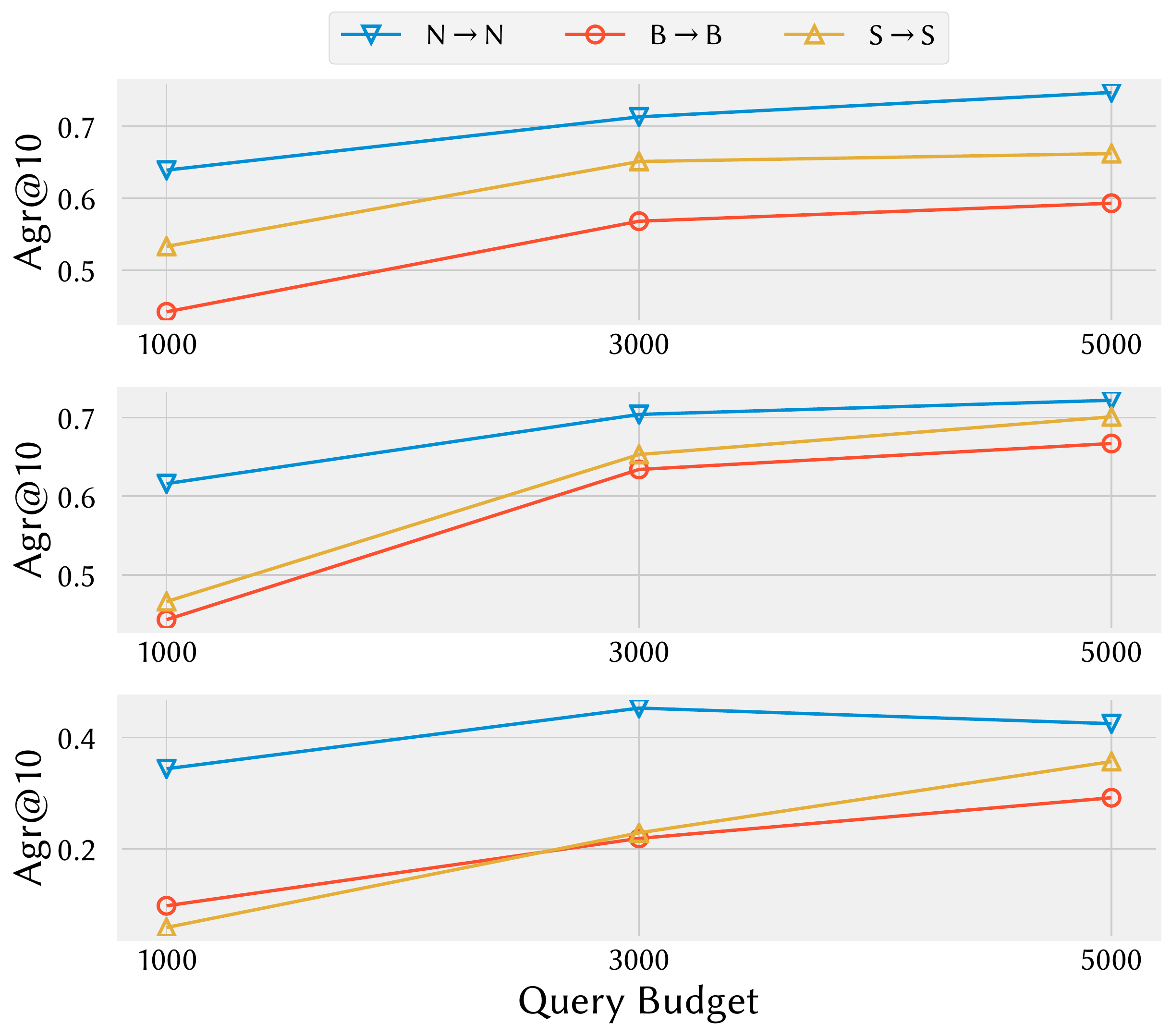}  
	\end{subfigure}
\caption{Influence of query budgets on ML-1M (top), Steam (middle) and Beauty (bottom).}
\label{fig:budget}
}
\vspace{-22pt}
\end{table}

\xhdr{Data Sparsity.} In the previous experiments, we notice that the sparsity of the original dataset on which the black-box is trained might influence the quality of our distilled model, suggested by the results on the Beauty dataset from \Cref{tab:distill}. For the sparsity problem, we base a series of experiments on the most sparse dataset (Beauty); all three models are used to study whether model extraction performance deterioration is related to the dataset. We choose slightly different preprocessing techniques to build a $k$-core Beauty dataset (i.e.~densify interactions until item / user frequencies are both $\ge k$). The processed $k$-core datasets are denser with increasing $k$. Our item candidate size (100 negatives) in evaluation does not change. Black-box models are trained on such processed data followed by autoregressive extraction experiments with a $5000$ sequence budget in \Cref{tab:distillation_sparsity}. As sparsity drops, compared to the 5-core Beauty data, both black-box and extracted models perform better, where the increasing Agr@10 indicates that the extracted models become more `similar' to the black-box model. Our results indicate that data sparisty is an important factor for model extraction performance, where training on denser 
%training 
datasets usually leads to stronger extracted models.

\xhdr{Budget.} We also want to find out how important the query budget is for distillation performance. 
%Since
In our framework we assume that the attacker can only query the black-box model a limited number of times (corresponding to a limited budget). Intuitively, a larger budget would induce a larger generated dataset for distillation and lead to a better distilled model with higher similarity and stronger recommendation scores. However, 
%query comes with certain prices and 
%it's desirable 
it is preferable to find an optimal budget such that the white-box system is `close enough' to generate adversarial examples and form threats. % Thus, we conduct more experiments 
%based 
% to find out what budget 
%would be the best fit 
% is sufficient for our purpose; see 
Our experiments in \Cref{fig:budget} suggest that an increasing budget would lead to rapid
initial improvements,
%in the first place,
resulting in a highly similar white-box model. 
%With budget increasing, 
Beyond a certain point,
the 
%gaining starts to decrease 
gains are marginal.

% \begin{table}[h]
% \begin{tabular}{@{}c|ccc|ccc@{}}
% \toprule
%       & \multicolumn{3}{c}{Random} & \multicolumn{3}{c}{Autoregressive} \\
% Budget & \multicolumn{1}{c}{KL-Div} & \multicolumn{1}{c}{Arg@10} & \multicolumn{1}{c}{R@10} & \multicolumn{1}{c}{KL-Div} & \multicolumn{1}{c}{Arg@10} & \multicolumn{1}{c}{R@10} \\
% \midrule
% 1k     & 0.326  & 0.406  & 0.774          & \textbf{0.102} & \textbf{0.684} & \textbf{0.783} \\
% 2k     & 0.315  & 0.418  & 0.786          & \textbf{0.076} & \textbf{0.733} & \textbf{0.797} \\
% 3k     & 0.296  & 0.455  & 0.791          & \textbf{0.067} & \textbf{0.750} & \textbf{0.803} \\
% 4k     & 0.212  & 0.584  & \textbf{0.810} & \textbf{0.039} & \textbf{0.823} & 0.809          \\
% 5k     & 0.207  & 0.593  & 0.809          & \textbf{0.032} & \textbf{0.839} & \textbf{0.810} \\
% \bottomrule
% \end{tabular}
% \caption{NARM on ML-1M distillation with different budgets}
% \label{tab:distillation_budget}
% \end{table}

\subsection{RQ3: Can We Perform Profile Pollution Attacks using the Extracted Model?}
\label{exp:rq3}

\begin{figure}[t]
  \begin{subfigure}{\textwidth}
    \centering
  % include first image
    \includegraphics[width=\linewidth]{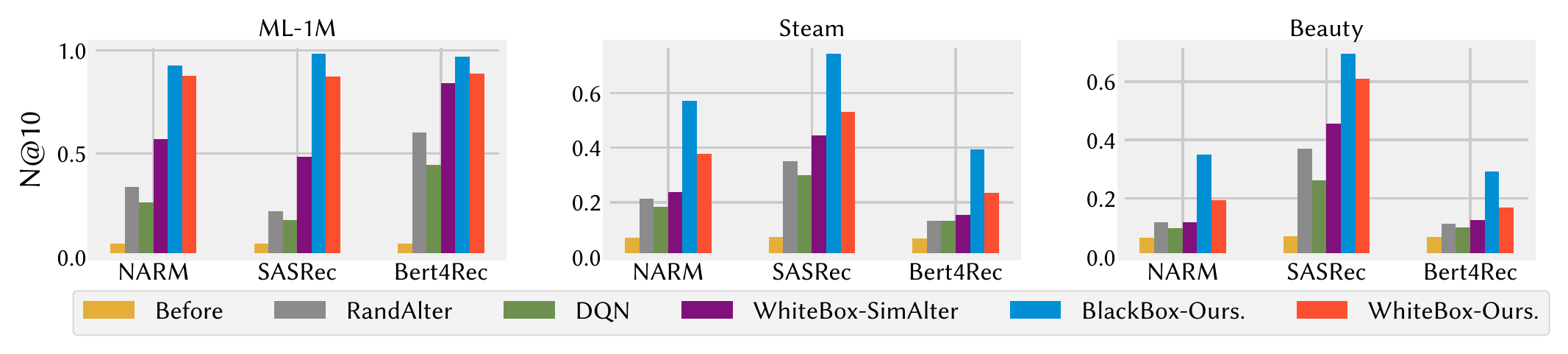}  
  \end{subfigure}
  \vspace{-10pt}
  \caption{Profile pollution attacks performance comparisons with different methods.}
  \vspace{-16pt}
  \label{fig:pollution}
\end{figure}

\xhdr{Setup.} In profile pollution, we inject adversarial items after original user history and test the corrupted data on the black-box recommender. The average lengths of ML-1M, Steam and Beauty are 166, 13 and 9 respectively;
%therefore, 
based on this
we generate $10$ adversarial items for ML-1M and $2$ items for Steam and Beauty. We perform profile pollution attacks on all users and present the average metrics in \Cref{fig:pollution}. We avoid repeating targets in injection items to rule out trivial solutions like sequence of solely target items. Based on this setting, we introduce the following baseline methods: (1) \emph{RandAlter}: alternating interactions with random items and target item(s)~\cite{song2020poisonrec}; (2) \emph{Deep Q Learning (DQN)}: naive Q learning model with RNN-based architecture, the rank and number of target item(s) in top-k recommendations are used as training rewards~\cite{zhao2017deep, zhang2020practical}; (3) \emph{WhiteBox SimAlter}: alternating interactions with similar items and target item(s), where similar items could be computed based on similarity of the white-box item embeddings. (4) In addition, we experiment 
%on our approach 
with
the black-box recommender as
a
surrogate model and perform our attacks (\emph{BlackBox-Ours.}).

% \todo{Attacks Rules + Baselines}

\xhdr{General Attack Performance Comparisons.} In \Cref{fig:pollution}, we present the profile pollution performance compared with baselines on three different datasets. (1)~Comparing our results with \emph{BlackBox-Ours.} black-box models show the  vulnerabilities to pollution attacks from extracted white-box model generally. We notice that the attacking performance of distilled models is comparable in ML-1M, but leads to worsening metrics as datasets become sparser and recommenders become harder to extract, for eaxmple, the average metrics of NARM on ML-1M reach 94.8\% attack performance of the black-box model, compared to 66.2\% on Steam and 55.6\% on Beauty. 
(2)~On all datasets, our method 
%performs 
achieves
the best targeted item promotion results. For example, on Steam,
%dataset, 
N@10 scores of 
%these 
the
targeted items are significantly improved from 0.070 to 0.381. 
%It is because our method 
This shows the benefit of
%exploit 
exploiting
the surrogate model from model extraction, and our attacking method is designed as an adversarial example method~\cite{goodfellow2014explaining}, which surpasses heuristic methods 
%without 
that lack
knowledge of the victim models.
(3)~Empricially, we find that the robustness varies for different victim recommenders. For example, results on Steam and Beauty datasets show SASRec is the most vulnerable model under attacks in our experiments. For instance, N@10 of SASRec increases from 0.071 to 0.571 
%averagely
on average
on Steam and Beauty datasets, but N@10 of NARM increases from 0.0680 to 0.286.

% The upper curve shows the curve of NARM, followed by SASRec and BERT4Rec. Surprisingly, the similarity between distilled and black-box recommenders seems to be less important in this case, as both 
%attacking 

% The upper curve presents the target exposure improvements on Recall@1 while the lower one on Recall@10, and the horizontal axis reflects the items' total appearances from high to low. We notice that popular items (left side of the figure) are generally more vulnerable against targeted attacks and could be easily manipulated for gaining attention. Interestingly, individual items have different degrees of robustness when facing such targeted attacks.

% \begin{figure}[h]
%     \centering
%     \includegraphics[width=1\linewidth]{paper/attack_example.png}
%     \caption{Adversarial Attack on Test Data}
%     \label{fig:attack_curve}
% \end{figure}

% In \Cref{fig:attack_curve}, rows represent white-box architectures in the order of NARM, SASRec and BERT4Rec, while three columns stand for three different datasets: ML-1M, Steam and Beauty. 

\begin{table}[t]
\small
\begin{tabular}{ccccccccccc}
\toprule
                        &        & \multicolumn{3}{c}{ML-1M}                        & \multicolumn{3}{c}{Steam}                        & \multicolumn{3}{c}{Beauty}                       \\ \cmidrule(l){3-5} \cmidrule(l){6-8}  \cmidrule(l){9-11} 
Popularity              & Attack & NARM           & SASRec         & BERT4Rec       & NARM           & SASRec         & BERT4Rec       & NARM           & SASRec         & BERT4Rec       \\ \midrule
\multirow{2}{*}{head}   & before & 0.202          & 0.217          & 0.201          & 0.313          & 0.327          & 0.311          & 0.261          & 0.246          & 0.260          \\
                        & ours   & \textbf{0.987} & \textbf{0.981} & \textbf{0.968} & \textbf{0.850} & \textbf{0.745} & \textbf{0.714} & \textbf{0.650} & \textbf{0.825} & \textbf{0.521} \\ \midrule
\multirow{2}{*}{middle} & before & 0.037          & 0.034          & 0.036          & 0.012          & 0.012          & 0.009          & 0.023          & 0.030          & 0.027          \\
                        & ours   & \textbf{0.902} & \textbf{0.876} & \textbf{0.901} & \textbf{0.341} & \textbf{0.585} & \textbf{0.152} & \textbf{0.106} & \textbf{0.556} & \textbf{0.105} \\ \midrule
\multirow{2}{*}{tail}   & before & 0.005          & 0.008          & 0.009          & 0.000          & 0.000          & \textbf{0.000} & \textbf{0.002} & 0.009          & 0.002          \\
                        & ours   & \textbf{0.701} & \textbf{0.760} & \textbf{0.760} & \textbf{0.017} & \textbf{0.160} & \textbf{0.000} & 0.001          & \textbf{0.557} & \textbf{0.010} \\ \bottomrule
\end{tabular}
\caption{Profile pollution attacks to different sequential models for items with different popularity, reported in N@10.}\vspace{-16pt}
\label{tab:pollution-pop}
\end{table}

\xhdr{Items w/ Different Popularities.} \Cref{tab:pollution-pop} shows our profile pollution attacks to items with different popularities. We group target items using 
%this 
the following
rules~\cite{anderson2006long}: \emph{head} denotes the top 20\%, \emph{tail} is the bottom 20\% and \emph{middle} is the rest according to the item appearance frequency (popularity). From~\Cref{tab:pollution-pop}, our attack method is effective for items with different popularities. But in all scenarios, ranking results after attacking  decrease as the popularity of the target item declines; popular items are generally more vulnerable under targeted attacks and could easily be manipulated for gaining attention. Unpopular items, however, are often harder to
%be attacked for more exposure.
attack.

\subsection{RQ4: Can We Perform Data Poisoning Attacks using the Extracted Model?}
\label{exp:rq4}

\begin{figure}[ht]
  \begin{subfigure}{\textwidth}
    \centering
  % include first image
    \includegraphics[width=\linewidth]{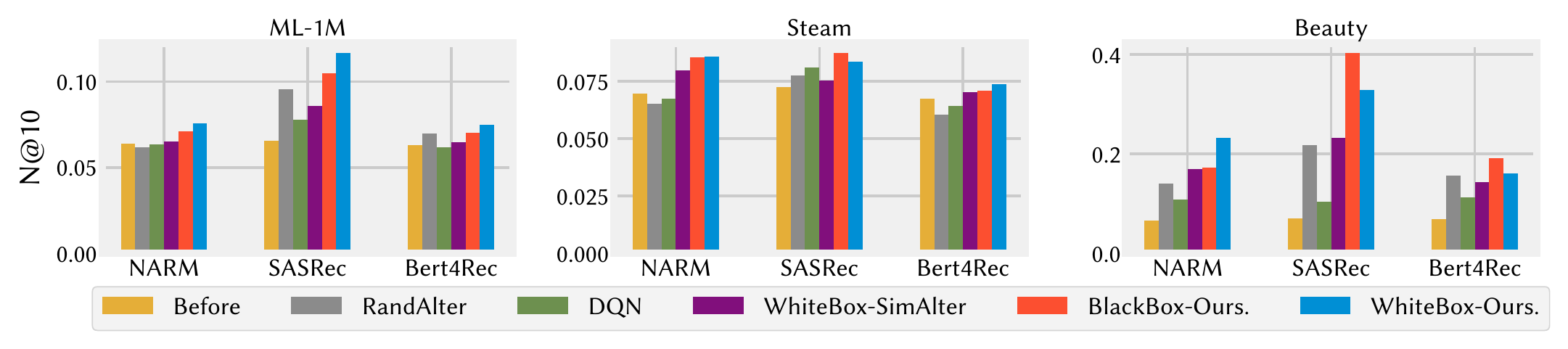}
  \end{subfigure}
  \vspace{-10pt}
  \caption{Data poisoning attacks performance comparisons with different methods}
  \vspace{-16pt}
  \label{fig:poisoning}
\end{figure}

\xhdr{Setup.} Different from profile pollution, we do not select a single item as a target and use target groups instead as the attack objective to avoid a large number of similar injection profiles and accelerate retraining. Target selection is identical to profile pollution and we randomly select an attack target from the 25 items in each step during the generation of adversarial profiles. Then, the black-box recommender is retrained once and tested on each of the target items, we present the average results in \Cref{fig:poisoning}. We follow~\cite{tang2020revisiting} to generate profiles equivalent to $1\%$ of 
%user number 
the number of users
from the original dataset and adopt the same baseline methods in profile pollution.

% \todo{Attack Rules}.

% $1000$ and adopt six different values of the original data size ${0, 250, 500, 1000, 2000, 4000}$. 
%In the case of $0$, 

% \begin{figure*}[!htbp]
% \centering
%     \begin{subfigure}[b]{0.33\textwidth}
%       \resizebox{\linewidth}{!}
%       {\input{paper/images/retrain/narm_retrain}}
%       \caption{NARM, ML-1M}
%      \end{subfigure}
%     \begin{subfigure}[b]{0.33\textwidth}
%       \resizebox{\linewidth}{!}
%       {\input{paper/images/retrain/sas_retrain}}
%       \caption{SASRec, ML-1M}
% 	\end{subfigure}
% 	\begin{subfigure}[b]{0.33\textwidth}
%       \resizebox{\linewidth}{!}
%       {\input{paper/images/retrain/bert_retrain}}
%       \caption{BERT4Rec, ML-1M}
% 	\end{subfigure}
% \caption{Data poisoning metrics of NARM, SASRec and BERT4Rec on ML-1M}
% \label{fig:attack_curve_retrain}
% \end{figure*}

\xhdr{General Attack Performance Comparisons.} (1) Our methods surpass other baselines but overall promotion is not as effective as profile pollution. It is because we adopt multiple targets for simultaneous poisoning, and profile pollution attacks specific user with profiles information, where attacking examples can be stronger. (2) Compared to \emph{RandAlter}, the proposed adversarial profile generation further enhances this advantage by connecting the unlikely popular items with targets and magnifying the bias for more exposure of our target items. For example in Beauty, the average N@10 is 0.066 before attack against 0.240 by our method across three models. 
%, with a significant N@10 improvement of 0.138 on average, compared to 0.055 for BERT4Rec and 0.061 for NARM. 
Moreover, we notice that \emph{DQN} performs 
%not as good as 
worse than
in profile pollution and could occasionally lead to performance deterioration. Potential reasons for the performance drop could be: Less frequent appearance of target items against the co-visitation approach; Updated model parameters are independent from the generated fake profiles, as in~\cite{tang2020revisiting}. (3) Comparable results between  \emph{BlackBox-Ours.} and \emph{WhiteBox-Ours.} suggest the bottleneck for data poisoning can be generation algorithm instead of white-box similarity.

\begin{table}[t]
\small

\begin{tabular}{ccccccccccc}
\toprule
                        &        & \multicolumn{3}{c}{ML-1M}                        & \multicolumn{3}{c}{Steam}                        & \multicolumn{3}{c}{Beauty}                       \\ \cmidrule(l){3-5} \cmidrule(l){6-8}  \cmidrule(l){9-11} 
Popularity              & Attack & NARM           & SASRec         & BERT4Rec       & NARM           & SASRec         & BERT4Rec       & NARM           & SASRec         & BERT4Rec       \\ \midrule
\multirow{2}{*}{head}   & before    & 0.202          & 0.217          & 0.201          & 0.313          & 0.327          & 0.311          & 0.261          & 0.246          & 0.260          \\
                        & ours      & \textbf{0.205} & \textbf{0.351} & \textbf{0.213} & \textbf{0.347} & \textbf{0.352} & \textbf{0.324} & \textbf{0.425} & \textbf{0.477} & \textbf{0.364} \\ \midrule
\multirow{2}{*}{medium} & before    & 0.037          & 0.034          & 0.036          & 0.012          & 0.012          & 0.009          & 0.023          & 0.030          & 0.027          \\
                        & ours      & \textbf{0.053} & \textbf{0.067} & \textbf{0.048} & \textbf{0.026} & \textbf{0.021} & \textbf{0.014} & \textbf{0.217} & \textbf{0.309} & \textbf{0.135} \\ \midrule
\multirow{2}{*}{tail}   & before    & 0.005          & 0.008          & 0.009          & 0.000          & 0.000          & 0.000          & 0.002          & 0.009          & 0.002          \\
                        & ours      & \textbf{0.016} & \textbf{0.032} & \textbf{0.017} & \textbf{0.004} & \textbf{0.004} & \textbf{0.004} & \textbf{0.086} & \textbf{0.235} & \textbf{0.036} \\ \bottomrule
\end{tabular}

\caption{Data poisoning attacks to different sequential models for items with different popularity, reported as N@10.}\vspace{-16pt}
\label{tab:poisoning-pop}
\end{table}

\xhdr{Items w/ Different Popularities.}
\Cref{tab:poisoning-pop} reveals the poisoning effectiveness 
%on different 
as a function of
target popularity. In contrast to the numbers in \Cref{tab:pollution-pop}, the relative improvements are more significant for \emph{middle} and \emph{tail} items. For example, the relative improvement for \emph{head} is 30.8\%, compared to over 300\% for \emph{middle} items and over 1000\% for \emph{tail} items in N@10. The results suggest that data poisoning is more helpful in elevating exposure for less popular items, while the promotion attacks of popular items via profile injection can be harder in this case.

\section{Conclusion And Future Work}
In this work, we systematically explore the feasibility and efficacy of stealing and attacking sequential recommender systems with different state-of-the-art architectures. First, our experimental results show that black-box models can be threatened by model extraction attacks. That is, we are able to learn a white-box model which behaves 
%highly 
similarly (for instance, with 0.747 top-$10$ agreement on the ML-1M dataset) to the black-box model without
%the
access to training data. This suggests that attacks on the white-box could be transferred to the black-box model. To verify this intuition, we conduct further experiments to study the vulnerability of black-box models using profile pollution and data poisoning attacks. Our experiments show that the performance of a well-trained black-box model can be drastically biased and corrupted under both attacks.

% (for instance, when attacking items using NARM, the probability of item exposure in the top-$10$ recommendations could rise up to $0.987$ in terms of NDCG@10).

For future work, first, we can extend our framework to more universal settings. In particular, how can we perform model extraction attacks for matrix factorization models or graph-based models? Second, it would be interesting to develop defense algorithms or other novel robust training pipelines, so that sequential recommender systems could be more robust against adversarial and data poisoning attacks. Third, 
%activate 
active
learning can be applied to find more effective sampling strategies for model extraction with fewer queries.

%%%%%%%%%%%%%%%%%%%%%%%%%%%%%%%%%%%%%%%%%%%%%%%%%%

\bibliographystyle{ACM-Reference-Format}
\bibliography{reference}


\begin{thebibliography}{46}

%%% the Web address.

\ifx \showCODEN    \undefined \def \showCODEN     #1{\unskip}     \fi
\ifx \showDOI      \undefined \def \showDOI       #1{#1}\fi
\ifx \showISBNx    \undefined \def \showISBNx     #1{\unskip}     \fi
\ifx \showISBNxiii \undefined \def \showISBNxiii  #1{\unskip}     \fi
\ifx \showISSN     \undefined \def \showISSN      #1{\unskip}     \fi
\ifx \showLCCN     \undefined \def \showLCCN      #1{\unskip}     \fi
\ifx \shownote     \undefined \def \shownote      #1{#1}          \fi
\ifx \showarticletitle \undefined \def \showarticletitle #1{#1}   \fi
\ifx \showURL      \undefined \def \showURL       {\relax}        \fi
% invisible to TeX
\providecommand\bibfield[2]{#2}
\providecommand\bibinfo[2]{#2}
\providecommand\natexlab[1]{#1}
\providecommand\showeprint[2][]{arXiv:#2}

\bibitem[\protect\citeauthoryear{Anderson}{Anderson}{2006}]%
        {anderson2006long}
\bibfield{author}{\bibinfo{person}{Chris Anderson}.}
  \bibinfo{year}{2006}\natexlab{}.
\newblock \bibinfo{booktitle}{\emph{The long tail: Why the future of business
  is selling less of more}}.
\newblock \bibinfo{publisher}{Hachette Books}.
\newblock


\bibitem[\protect\citeauthoryear{Burke, Mobasher, and Bhaumik}{Burke
  et~al\mbox{.}}{2005}]%
        {burke2005limited}
\bibfield{author}{\bibinfo{person}{Robin Burke}, \bibinfo{person}{Bamshad
  Mobasher}, {and} \bibinfo{person}{Runa Bhaumik}.}
  \bibinfo{year}{2005}\natexlab{}.
\newblock \showarticletitle{Limited knowledge shilling attacks in collaborative
  filtering systems}. In \bibinfo{booktitle}{\emph{Proceedings of 3rd
  international workshop on intelligent techniques for web personalization
  (ITWP 2005), 19th international joint conference on artificial intelligence
  (IJCAI 2005)}}. \bibinfo{pages}{17--24}.
\newblock


\bibitem[\protect\citeauthoryear{Cho, van Merrienboer, G{\"u}l{\c{c}}ehre,
  Bahdanau, Bougares, Schwenk, and Bengio}{Cho et~al\mbox{.}}{2014}]%
        {cho2014learning}
\bibfield{author}{\bibinfo{person}{Kyunghyun Cho}, \bibinfo{person}{Bart van
  Merrienboer}, \bibinfo{person}{{\c{C}}aglar G{\"u}l{\c{c}}ehre},
  \bibinfo{person}{Dzmitry Bahdanau}, \bibinfo{person}{Fethi Bougares},
  \bibinfo{person}{Holger Schwenk}, {and} \bibinfo{person}{Yoshua Bengio}.}
  \bibinfo{year}{2014}\natexlab{}.
\newblock \showarticletitle{Learning Phrase Representations using RNN
  Encoder-Decoder for Statistical Machine Translation}. In
  \bibinfo{booktitle}{\emph{EMNLP}}.
\newblock


\bibitem[\protect\citeauthoryear{Christakopoulou and Banerjee}{Christakopoulou
  and Banerjee}{2019}]%
        {christakopoulou2019adversarial}
\bibfield{author}{\bibinfo{person}{Konstantina Christakopoulou} {and}
  \bibinfo{person}{Arindam Banerjee}.} \bibinfo{year}{2019}\natexlab{}.
\newblock \showarticletitle{Adversarial attacks on an oblivious recommender}.
  In \bibinfo{booktitle}{\emph{Proceedings of the 13th ACM Conference on
  Recommender Systems}}. \bibinfo{pages}{322--330}.
\newblock


\bibitem[\protect\citeauthoryear{Devlin, Chang, Lee, and Toutanova}{Devlin
  et~al\mbox{.}}{2018}]%
        {jacob2018bert}
\bibfield{author}{\bibinfo{person}{Jacob Devlin}, \bibinfo{person}{Ming-Wei
  Chang}, \bibinfo{person}{Kenton Lee}, {and} \bibinfo{person}{Kristina~N.
  Toutanova}.} \bibinfo{year}{2018}\natexlab{}.
\newblock \showarticletitle{BERT: Pre-training of Deep Bidirectional
  Transformers for Language Understanding}.
\newblock
\urldef\tempurl%
\url{https://arxiv.org/abs/1810.04805}
\showURL{%
\tempurl}


\bibitem[\protect\citeauthoryear{Fang, Gong, and Liu}{Fang
  et~al\mbox{.}}{2020}]%
        {fang2020influence}
\bibfield{author}{\bibinfo{person}{Minghong Fang},
  \bibinfo{person}{Neil~Zhenqiang Gong}, {and} \bibinfo{person}{Jia Liu}.}
  \bibinfo{year}{2020}\natexlab{}.
\newblock \showarticletitle{Influence function based data poisoning attacks to
  top-n recommender systems}. In \bibinfo{booktitle}{\emph{Proceedings of The
  Web Conference 2020}}. \bibinfo{pages}{3019--3025}.
\newblock


\bibitem[\protect\citeauthoryear{Goodfellow, Pouget-Abadie, Mirza, Xu,
  Warde-Farley, Ozair, Courville, and Bengio}{Goodfellow
  et~al\mbox{.}}{2014a}]%
        {goodfellow2014generative}
\bibfield{author}{\bibinfo{person}{Ian Goodfellow}, \bibinfo{person}{Jean
  Pouget-Abadie}, \bibinfo{person}{Mehdi Mirza}, \bibinfo{person}{Bing Xu},
  \bibinfo{person}{David Warde-Farley}, \bibinfo{person}{Sherjil Ozair},
  \bibinfo{person}{Aaron Courville}, {and} \bibinfo{person}{Yoshua Bengio}.}
  \bibinfo{year}{2014}\natexlab{a}.
\newblock \showarticletitle{Generative adversarial nets}. In
  \bibinfo{booktitle}{\emph{Advances in neural information processing
  systems}}. \bibinfo{pages}{2672--2680}.
\newblock


\bibitem[\protect\citeauthoryear{Goodfellow, Shlens, and Szegedy}{Goodfellow
  et~al\mbox{.}}{2014b}]%
        {goodfellow2014explaining}
\bibfield{author}{\bibinfo{person}{Ian~J Goodfellow}, \bibinfo{person}{Jonathon
  Shlens}, {and} \bibinfo{person}{Christian Szegedy}.}
  \bibinfo{year}{2014}\natexlab{b}.
\newblock \showarticletitle{Explaining and harnessing adversarial examples}.
\newblock \bibinfo{journal}{\emph{arXiv preprint arXiv:1412.6572}}
  (\bibinfo{year}{2014}).
\newblock


\bibitem[\protect\citeauthoryear{Gunes, Kaleli, Bilge, and Polat}{Gunes
  et~al\mbox{.}}{2014}]%
        {gunes2014shilling}
\bibfield{author}{\bibinfo{person}{Ihsan Gunes}, \bibinfo{person}{Cihan
  Kaleli}, \bibinfo{person}{Alper Bilge}, {and} \bibinfo{person}{Huseyin
  Polat}.} \bibinfo{year}{2014}\natexlab{}.
\newblock \showarticletitle{Shilling attacks against recommender systems: a
  comprehensive survey}.
\newblock \bibinfo{journal}{\emph{Artificial Intelligence Review}}
  \bibinfo{volume}{42}, \bibinfo{number}{4} (\bibinfo{year}{2014}),
  \bibinfo{pages}{767--799}.
\newblock


\bibitem[\protect\citeauthoryear{Harper and Konstan}{Harper and
  Konstan}{2015}]%
        {harper2015movielens}
\bibfield{author}{\bibinfo{person}{F~Maxwell Harper} {and}
  \bibinfo{person}{Joseph~A Konstan}.} \bibinfo{year}{2015}\natexlab{}.
\newblock \showarticletitle{The movielens datasets: History and context}.
\newblock \bibinfo{journal}{\emph{Acm transactions on interactive intelligent
  systems (tiis)}} \bibinfo{volume}{5}, \bibinfo{number}{4}
  (\bibinfo{year}{2015}), \bibinfo{pages}{1--19}.
\newblock


\bibitem[\protect\citeauthoryear{He and McAuley}{He and McAuley}{2016}]%
        {he2016fusing}
\bibfield{author}{\bibinfo{person}{Ruining He} {and} \bibinfo{person}{Julian
  McAuley}.} \bibinfo{year}{2016}\natexlab{}.
\newblock \showarticletitle{Fusing similarity models with markov chains for
  sparse sequential recommendation}. In \bibinfo{booktitle}{\emph{2016 IEEE
  16th International Conference on Data Mining (ICDM)}}. IEEE,
  \bibinfo{pages}{191--200}.
\newblock


\bibitem[\protect\citeauthoryear{He, Liao, Zhang, Nie, Hu, and Chua}{He
  et~al\mbox{.}}{2017}]%
        {he2017neural}
\bibfield{author}{\bibinfo{person}{Xiangnan He}, \bibinfo{person}{Lizi Liao},
  \bibinfo{person}{Hanwang Zhang}, \bibinfo{person}{Liqiang Nie},
  \bibinfo{person}{Xia Hu}, {and} \bibinfo{person}{Tat-Seng Chua}.}
  \bibinfo{year}{2017}\natexlab{}.
\newblock \showarticletitle{Neural collaborative filtering}. In
  \bibinfo{booktitle}{\emph{Proceedings of the 26th international conference on
  world wide web}}. \bibinfo{pages}{173--182}.
\newblock


\bibitem[\protect\citeauthoryear{Hidasi, Karatzoglou, Baltrunas, and
  Tikk}{Hidasi et~al\mbox{.}}{2015}]%
        {hidasi2015session}
\bibfield{author}{\bibinfo{person}{Bal{\'a}zs Hidasi},
  \bibinfo{person}{Alexandros Karatzoglou}, \bibinfo{person}{Linas Baltrunas},
  {and} \bibinfo{person}{Domonkos Tikk}.} \bibinfo{year}{2015}\natexlab{}.
\newblock \showarticletitle{Session-based recommendations with recurrent neural
  networks}.
\newblock \bibinfo{journal}{\emph{arXiv preprint arXiv:1511.06939}}
  (\bibinfo{year}{2015}).
\newblock


\bibitem[\protect\citeauthoryear{Hinton, Vinyals, and Dean}{Hinton
  et~al\mbox{.}}{2015}]%
        {hinton2015distilling}
\bibfield{author}{\bibinfo{person}{Geoffrey Hinton}, \bibinfo{person}{Oriol
  Vinyals}, {and} \bibinfo{person}{Jeff Dean}.}
  \bibinfo{year}{2015}\natexlab{}.
\newblock \showarticletitle{Distilling the knowledge in a neural network}.
\newblock \bibinfo{journal}{\emph{arXiv preprint arXiv:1503.02531}}
  (\bibinfo{year}{2015}).
\newblock


\bibitem[\protect\citeauthoryear{Huang, Mu, Gong, Li, Liu, and Xu}{Huang
  et~al\mbox{.}}{2021}]%
        {huang2021data}
\bibfield{author}{\bibinfo{person}{Hai Huang}, \bibinfo{person}{Jiaming Mu},
  \bibinfo{person}{Neil~Zhenqiang Gong}, \bibinfo{person}{Qi Li},
  \bibinfo{person}{Bin Liu}, {and} \bibinfo{person}{Mingwei Xu}.}
  \bibinfo{year}{2021}\natexlab{}.
\newblock \showarticletitle{Data Poisoning Attacks to Deep Learning Based
  Recommender Systems}.
\newblock \bibinfo{journal}{\emph{arXiv preprint arXiv:2101.02644}}
  (\bibinfo{year}{2021}).
\newblock


\bibitem[\protect\citeauthoryear{Jin, Jin, Zhou, and Szolovits}{Jin
  et~al\mbox{.}}{2019}]%
        {jin2019bert}
\bibfield{author}{\bibinfo{person}{Di Jin}, \bibinfo{person}{Zhijing Jin},
  \bibinfo{person}{Joey~Tianyi Zhou}, {and} \bibinfo{person}{Peter Szolovits}.}
  \bibinfo{year}{2019}\natexlab{}.
\newblock \showarticletitle{Is BERT Really Robust}.
\newblock \bibinfo{journal}{\emph{A Strong Baseline for Natural Language Attack
  on Text Classification and Entailment}} (\bibinfo{year}{2019}).
\newblock


\bibitem[\protect\citeauthoryear{Kang and McAuley}{Kang and McAuley}{2018}]%
        {kang2018self}
\bibfield{author}{\bibinfo{person}{Wang-Cheng Kang} {and}
  \bibinfo{person}{Julian McAuley}.} \bibinfo{year}{2018}\natexlab{}.
\newblock \showarticletitle{Self-attentive sequential recommendation}. In
  \bibinfo{booktitle}{\emph{2018 IEEE International Conference on Data Mining
  (ICDM)}}. IEEE, \bibinfo{pages}{197--206}.
\newblock


\bibitem[\protect\citeauthoryear{Kariyappa, Prakash, and Qureshi}{Kariyappa
  et~al\mbox{.}}{2020}]%
        {kariyappa2020maze}
\bibfield{author}{\bibinfo{person}{Sanjay Kariyappa}, \bibinfo{person}{Atul
  Prakash}, {and} \bibinfo{person}{Moinuddin Qureshi}.}
  \bibinfo{year}{2020}\natexlab{}.
\newblock \showarticletitle{MAZE: Data-Free Model Stealing Attack Using
  Zeroth-Order Gradient Estimation}.
\newblock \bibinfo{journal}{\emph{arXiv preprint arXiv:2005.03161}}
  (\bibinfo{year}{2020}).
\newblock


\bibitem[\protect\citeauthoryear{Kingma and Ba}{Kingma and Ba}{2014}]%
        {kingma2014adam}
\bibfield{author}{\bibinfo{person}{Diederik~P Kingma} {and}
  \bibinfo{person}{Jimmy Ba}.} \bibinfo{year}{2014}\natexlab{}.
\newblock \showarticletitle{Adam: A method for stochastic optimization}.
\newblock \bibinfo{journal}{\emph{arXiv preprint arXiv:1412.6980}}
  (\bibinfo{year}{2014}).
\newblock


\bibitem[\protect\citeauthoryear{Krishna, Tomar, Parikh, Papernot, and
  Iyyer}{Krishna et~al\mbox{.}}{2020}]%
        {Krishna2020Thieves}
\bibfield{author}{\bibinfo{person}{Kalpesh Krishna},
  \bibinfo{person}{Gaurav~Singh Tomar}, \bibinfo{person}{Ankur~P. Parikh},
  \bibinfo{person}{Nicolas Papernot}, {and} \bibinfo{person}{Mohit Iyyer}.}
  \bibinfo{year}{2020}\natexlab{}.
\newblock \showarticletitle{Thieves on Sesame Street! Model Extraction of
  BERT-based APIs}. In \bibinfo{booktitle}{\emph{International Conference on
  Learning Representations}}.
\newblock
\urldef\tempurl%
\url{https://openreview.net/forum?id=Byl5NREFDr}
\showURL{%
\tempurl}


\bibitem[\protect\citeauthoryear{Kurakin, Goodfellow, and Bengio}{Kurakin
  et~al\mbox{.}}{2016}]%
        {kurakin2016adversarial}
\bibfield{author}{\bibinfo{person}{Alexey Kurakin}, \bibinfo{person}{Ian
  Goodfellow}, {and} \bibinfo{person}{Samy Bengio}.}
  \bibinfo{year}{2016}\natexlab{}.
\newblock \showarticletitle{Adversarial examples in the physical world}.
\newblock \bibinfo{journal}{\emph{arXiv preprint arXiv:1607.02533}}
  (\bibinfo{year}{2016}).
\newblock


\bibitem[\protect\citeauthoryear{Lam and Riedl}{Lam and Riedl}{2004}]%
        {lam2004shilling}
\bibfield{author}{\bibinfo{person}{Shyong~K Lam} {and} \bibinfo{person}{John
  Riedl}.} \bibinfo{year}{2004}\natexlab{}.
\newblock \showarticletitle{Shilling recommender systems for fun and profit}.
  In \bibinfo{booktitle}{\emph{Proceedings of the 13th international conference
  on World Wide Web}}. \bibinfo{pages}{393--402}.
\newblock


\bibitem[\protect\citeauthoryear{Lee, Hwang, and Ryu}{Lee
  et~al\mbox{.}}{2017}]%
        {lee2017all}
\bibfield{author}{\bibinfo{person}{Sungho Lee}, \bibinfo{person}{Sungjae
  Hwang}, {and} \bibinfo{person}{Sukyoung Ryu}.}
  \bibinfo{year}{2017}\natexlab{}.
\newblock \showarticletitle{All about activity injection: Threats, semantics,
  and detection}. In \bibinfo{booktitle}{\emph{2017 32nd IEEE/ACM International
  Conference on Automated Software Engineering (ASE)}}. IEEE,
  \bibinfo{pages}{252--262}.
\newblock


\bibitem[\protect\citeauthoryear{Li, Wang, Singh, and Vorobeychik}{Li
  et~al\mbox{.}}{2016}]%
        {li2016data}
\bibfield{author}{\bibinfo{person}{Bo Li}, \bibinfo{person}{Yining Wang},
  \bibinfo{person}{Aarti Singh}, {and} \bibinfo{person}{Yevgeniy Vorobeychik}.}
  \bibinfo{year}{2016}\natexlab{}.
\newblock \showarticletitle{Data poisoning attacks on factorization-based
  collaborative filtering}.
\newblock \bibinfo{journal}{\emph{arXiv preprint arXiv:1608.08182}}
  (\bibinfo{year}{2016}).
\newblock


\bibitem[\protect\citeauthoryear{Li, Ren, Chen, Ren, Lian, and Ma}{Li
  et~al\mbox{.}}{2017}]%
        {li2017neural}
\bibfield{author}{\bibinfo{person}{Jing Li}, \bibinfo{person}{Pengjie Ren},
  \bibinfo{person}{Zhumin Chen}, \bibinfo{person}{Zhaochun Ren},
  \bibinfo{person}{Tao Lian}, {and} \bibinfo{person}{Jun Ma}.}
  \bibinfo{year}{2017}\natexlab{}.
\newblock \showarticletitle{Neural attentive session-based recommendation}. In
  \bibinfo{booktitle}{\emph{Proceedings of the 2017 ACM on Conference on
  Information and Knowledge Management}}. \bibinfo{pages}{1419--1428}.
\newblock


\bibitem[\protect\citeauthoryear{Lowd and Meek}{Lowd and Meek}{2005}]%
        {lowd2005adversarial}
\bibfield{author}{\bibinfo{person}{Daniel Lowd} {and}
  \bibinfo{person}{Christopher Meek}.} \bibinfo{year}{2005}\natexlab{}.
\newblock \showarticletitle{Adversarial learning}. In
  \bibinfo{booktitle}{\emph{Proceedings of the eleventh ACM SIGKDD
  international conference on Knowledge discovery in data mining}}.
  \bibinfo{pages}{641--647}.
\newblock


\bibitem[\protect\citeauthoryear{McAuley, Targett, Shi, and Van
  Den~Hengel}{McAuley et~al\mbox{.}}{2015}]%
        {mcauley2015image}
\bibfield{author}{\bibinfo{person}{Julian McAuley},
  \bibinfo{person}{Christopher Targett}, \bibinfo{person}{Qinfeng Shi}, {and}
  \bibinfo{person}{Anton Van Den~Hengel}.} \bibinfo{year}{2015}\natexlab{}.
\newblock \showarticletitle{Image-based recommendations on styles and
  substitutes}. In \bibinfo{booktitle}{\emph{Proceedings of the 38th
  international ACM SIGIR conference on research and development in information
  retrieval}}. \bibinfo{pages}{43--52}.
\newblock


\bibitem[\protect\citeauthoryear{Merity, Xiong, Bradbury, and Socher}{Merity
  et~al\mbox{.}}{2016}]%
        {merity2016pointer}
\bibfield{author}{\bibinfo{person}{Stephen Merity}, \bibinfo{person}{Caiming
  Xiong}, \bibinfo{person}{James Bradbury}, {and} \bibinfo{person}{Richard
  Socher}.} \bibinfo{year}{2016}\natexlab{}.
\newblock \showarticletitle{Pointer sentinel mixture models}.
\newblock \bibinfo{journal}{\emph{arXiv preprint arXiv:1609.07843}}
  (\bibinfo{year}{2016}).
\newblock


\bibitem[\protect\citeauthoryear{Ni, Li, and McAuley}{Ni et~al\mbox{.}}{2019}]%
        {ni-etal-2019-justifying}
\bibfield{author}{\bibinfo{person}{Jianmo Ni}, \bibinfo{person}{Jiacheng Li},
  {and} \bibinfo{person}{Julian McAuley}.} \bibinfo{year}{2019}\natexlab{}.
\newblock \showarticletitle{Justifying Recommendations using Distantly-Labeled
  Reviews and Fine-Grained Aspects}. In \bibinfo{booktitle}{\emph{Proceedings
  of the 2019 Conference on Empirical Methods in Natural Language Processing
  and the 9th International Joint Conference on Natural Language Processing
  (EMNLP-IJCNLP)}}. \bibinfo{publisher}{Association for Computational
  Linguistics}, \bibinfo{address}{Hong Kong, China}, \bibinfo{pages}{188--197}.
\newblock
\urldef\tempurl%
\url{https://doi.org/10.18653/v1/D19-1018}
\showDOI{\tempurl}


\bibitem[\protect\citeauthoryear{Orekondy, Schiele, and Fritz}{Orekondy
  et~al\mbox{.}}{2019}]%
        {orekondy2019knockoff}
\bibfield{author}{\bibinfo{person}{Tribhuvanesh Orekondy},
  \bibinfo{person}{Bernt Schiele}, {and} \bibinfo{person}{Mario Fritz}.}
  \bibinfo{year}{2019}\natexlab{}.
\newblock \showarticletitle{Knockoff nets: Stealing functionality of black-box
  models}. In \bibinfo{booktitle}{\emph{Proceedings of the IEEE Conference on
  Computer Vision and Pattern Recognition}}. \bibinfo{pages}{4954--4963}.
\newblock


\bibitem[\protect\citeauthoryear{Pal, Gupta, Shukla, Kanade, Shevade, and
  Ganapathy}{Pal et~al\mbox{.}}{2019}]%
        {pal2019framework}
\bibfield{author}{\bibinfo{person}{Soham Pal}, \bibinfo{person}{Yash Gupta},
  \bibinfo{person}{Aditya Shukla}, \bibinfo{person}{Aditya Kanade},
  \bibinfo{person}{Shirish Shevade}, {and} \bibinfo{person}{Vinod Ganapathy}.}
  \bibinfo{year}{2019}\natexlab{}.
\newblock \showarticletitle{A framework for the extraction of deep neural
  networks by leveraging public data}.
\newblock \bibinfo{journal}{\emph{arXiv preprint arXiv:1905.09165}}
  (\bibinfo{year}{2019}).
\newblock


\bibitem[\protect\citeauthoryear{Papernot, McDaniel, Goodfellow, Jha, Celik,
  and Swami}{Papernot et~al\mbox{.}}{2017}]%
        {papernot2017practical}
\bibfield{author}{\bibinfo{person}{Nicolas Papernot}, \bibinfo{person}{Patrick
  McDaniel}, \bibinfo{person}{Ian Goodfellow}, \bibinfo{person}{Somesh Jha},
  \bibinfo{person}{Z~Berkay Celik}, {and} \bibinfo{person}{Ananthram Swami}.}
  \bibinfo{year}{2017}\natexlab{}.
\newblock \showarticletitle{Practical black-box attacks against machine
  learning}. In \bibinfo{booktitle}{\emph{Proceedings of the 2017 ACM on Asia
  conference on computer and communications security}}.
  \bibinfo{pages}{506--519}.
\newblock


\bibitem[\protect\citeauthoryear{Ren, Zhang, Xue, Wei, and Liu}{Ren
  et~al\mbox{.}}{2015}]%
        {ren2015towards}
\bibfield{author}{\bibinfo{person}{Chuangang Ren}, \bibinfo{person}{Yulong
  Zhang}, \bibinfo{person}{Hui Xue}, \bibinfo{person}{Tao Wei}, {and}
  \bibinfo{person}{Peng Liu}.} \bibinfo{year}{2015}\natexlab{}.
\newblock \showarticletitle{Towards discovering and understanding task
  hijacking in android}. In \bibinfo{booktitle}{\emph{24th $\{$USENIX$\}$
  Security Symposium ($\{$USENIX$\}$ Security 15)}}. \bibinfo{pages}{945--959}.
\newblock


\bibitem[\protect\citeauthoryear{Rendle, Freudenthaler, Gantner, and
  Schmidt-Thieme}{Rendle et~al\mbox{.}}{2012}]%
        {rendle2012bpr}
\bibfield{author}{\bibinfo{person}{Steffen Rendle}, \bibinfo{person}{Christoph
  Freudenthaler}, \bibinfo{person}{Zeno Gantner}, {and} \bibinfo{person}{Lars
  Schmidt-Thieme}.} \bibinfo{year}{2012}\natexlab{}.
\newblock \showarticletitle{BPR: Bayesian personalized ranking from implicit
  feedback}.
\newblock \bibinfo{journal}{\emph{arXiv preprint arXiv:1205.2618}}
  (\bibinfo{year}{2012}).
\newblock


\bibitem[\protect\citeauthoryear{Rendle, Freudenthaler, and
  Schmidt-Thieme}{Rendle et~al\mbox{.}}{2010}]%
        {rendle2010factorizing}
\bibfield{author}{\bibinfo{person}{Steffen Rendle}, \bibinfo{person}{Christoph
  Freudenthaler}, {and} \bibinfo{person}{Lars Schmidt-Thieme}.}
  \bibinfo{year}{2010}\natexlab{}.
\newblock \showarticletitle{Factorizing personalized markov chains for
  next-basket recommendation}. In \bibinfo{booktitle}{\emph{Proceedings of the
  19th international conference on World wide web}}. \bibinfo{pages}{811--820}.
\newblock


\bibitem[\protect\citeauthoryear{Song, Li, Hu, Wu, Li, Li, and Gao}{Song
  et~al\mbox{.}}{2020}]%
        {song2020poisonrec}
\bibfield{author}{\bibinfo{person}{Junshuai Song}, \bibinfo{person}{Zhao Li},
  \bibinfo{person}{Zehong Hu}, \bibinfo{person}{Yucheng Wu},
  \bibinfo{person}{Zhenpeng Li}, \bibinfo{person}{Jian Li}, {and}
  \bibinfo{person}{Jun Gao}.} \bibinfo{year}{2020}\natexlab{}.
\newblock \showarticletitle{Poisonrec: an adaptive data poisoning framework for
  attacking black-box recommender systems}. In \bibinfo{booktitle}{\emph{2020
  IEEE 36th International Conference on Data Engineering (ICDE)}}. IEEE,
  \bibinfo{pages}{157--168}.
\newblock


\bibitem[\protect\citeauthoryear{Sun, Liu, Wu, Pei, Lin, Ou, and Jiang}{Sun
  et~al\mbox{.}}{2019}]%
        {sun2019bert4rec}
\bibfield{author}{\bibinfo{person}{Fei Sun}, \bibinfo{person}{Jun Liu},
  \bibinfo{person}{Jian Wu}, \bibinfo{person}{Changhua Pei},
  \bibinfo{person}{Xiao Lin}, \bibinfo{person}{Wenwu Ou}, {and}
  \bibinfo{person}{Peng Jiang}.} \bibinfo{year}{2019}\natexlab{}.
\newblock \showarticletitle{BERT4Rec: Sequential recommendation with
  bidirectional encoder representations from transformer}. In
  \bibinfo{booktitle}{\emph{Proceedings of the 28th ACM International
  Conference on Information and Knowledge Management}}.
  \bibinfo{pages}{1441--1450}.
\newblock


\bibitem[\protect\citeauthoryear{Tang and Wang}{Tang and Wang}{2018}]%
        {tang2018personalized}
\bibfield{author}{\bibinfo{person}{Jiaxi Tang} {and} \bibinfo{person}{Ke
  Wang}.} \bibinfo{year}{2018}\natexlab{}.
\newblock \showarticletitle{Personalized top-n sequential recommendation via
  convolutional sequence embedding}. In \bibinfo{booktitle}{\emph{Proceedings
  of the Eleventh ACM International Conference on Web Search and Data Mining}}.
  \bibinfo{pages}{565--573}.
\newblock


\bibitem[\protect\citeauthoryear{Tang, Wen, and Wang}{Tang
  et~al\mbox{.}}{2020}]%
        {tang2020revisiting}
\bibfield{author}{\bibinfo{person}{Jiaxi Tang}, \bibinfo{person}{Hongyi Wen},
  {and} \bibinfo{person}{Ke Wang}.} \bibinfo{year}{2020}\natexlab{}.
\newblock \showarticletitle{Revisiting Adversarially Learned Injection Attacks
  Against Recommender Systems}. In \bibinfo{booktitle}{\emph{Fourteenth ACM
  Conference on Recommender Systems}}. \bibinfo{pages}{318--327}.
\newblock


\bibitem[\protect\citeauthoryear{Tram{\`e}r, Zhang, Juels, Reiter, and
  Ristenpart}{Tram{\`e}r et~al\mbox{.}}{2016}]%
        {tramer2016stealing}
\bibfield{author}{\bibinfo{person}{Florian Tram{\`e}r}, \bibinfo{person}{Fan
  Zhang}, \bibinfo{person}{Ari Juels}, \bibinfo{person}{Michael~K Reiter},
  {and} \bibinfo{person}{Thomas Ristenpart}.} \bibinfo{year}{2016}\natexlab{}.
\newblock \showarticletitle{Stealing machine learning models via prediction
  apis}. In \bibinfo{booktitle}{\emph{25th $\{$USENIX$\}$ Security Symposium
  ($\{$USENIX$\}$ Security 16)}}. \bibinfo{pages}{601--618}.
\newblock


\bibitem[\protect\citeauthoryear{Xing, Meng, Doozan, Snoeren, Feamster, and
  Lee}{Xing et~al\mbox{.}}{2013}]%
        {xing2013take}
\bibfield{author}{\bibinfo{person}{Xingyu Xing}, \bibinfo{person}{Wei Meng},
  \bibinfo{person}{Dan Doozan}, \bibinfo{person}{Alex~C Snoeren},
  \bibinfo{person}{Nick Feamster}, {and} \bibinfo{person}{Wenke Lee}.}
  \bibinfo{year}{2013}\natexlab{}.
\newblock \showarticletitle{Take this personally: Pollution attacks on
  personalized services}. In \bibinfo{booktitle}{\emph{22nd $\{$USENIX$\}$
  Security Symposium ($\{$USENIX$\}$ Security 13)}}. \bibinfo{pages}{671--686}.
\newblock


\bibitem[\protect\citeauthoryear{Yang, Gong, and Cai}{Yang
  et~al\mbox{.}}{2017}]%
        {yang2017fake}
\bibfield{author}{\bibinfo{person}{Guolei Yang},
  \bibinfo{person}{Neil~Zhenqiang Gong}, {and} \bibinfo{person}{Ying Cai}.}
  \bibinfo{year}{2017}\natexlab{}.
\newblock \showarticletitle{Fake Co-visitation Injection Attacks to Recommender
  Systems.}. In \bibinfo{booktitle}{\emph{NDSS}}.
\newblock


\bibitem[\protect\citeauthoryear{Zeller and Felten}{Zeller and Felten}{2008}]%
        {zeller2008cross}
\bibfield{author}{\bibinfo{person}{William Zeller} {and}
  \bibinfo{person}{Edward~W Felten}.} \bibinfo{year}{2008}\natexlab{}.
\newblock \showarticletitle{Cross-site request forgeries: Exploitation and
  prevention}.
\newblock \bibinfo{journal}{\emph{Bericht, Princeton University}}
  (\bibinfo{year}{2008}).
\newblock


\bibitem[\protect\citeauthoryear{Zhang, Li, Ding, and Gao}{Zhang
  et~al\mbox{.}}{2020}]%
        {zhang2020practical}
\bibfield{author}{\bibinfo{person}{Hengtong Zhang}, \bibinfo{person}{Yaliang
  Li}, \bibinfo{person}{Bolin Ding}, {and} \bibinfo{person}{Jing Gao}.}
  \bibinfo{year}{2020}\natexlab{}.
\newblock \showarticletitle{Practical Data Poisoning Attack against Next-Item
  Recommendation}. In \bibinfo{booktitle}{\emph{Proceedings of The Web
  Conference 2020}}. \bibinfo{pages}{2458--2464}.
\newblock


\bibitem[\protect\citeauthoryear{Zhao, Zhang, Xia, Ding, Yin, and Tang}{Zhao
  et~al\mbox{.}}{2017}]%
        {zhao2017deep}
\bibfield{author}{\bibinfo{person}{Xiangyu Zhao}, \bibinfo{person}{Liang
  Zhang}, \bibinfo{person}{Long Xia}, \bibinfo{person}{Zhuoye Ding},
  \bibinfo{person}{Dawei Yin}, {and} \bibinfo{person}{Jiliang Tang}.}
  \bibinfo{year}{2017}\natexlab{}.
\newblock \showarticletitle{Deep reinforcement learning for list-wise
  recommendations}.
\newblock \bibinfo{journal}{\emph{arXiv preprint arXiv:1801.00209}}
  (\bibinfo{year}{2017}).
\newblock


\bibitem[\protect\citeauthoryear{Zhou, Wu, Liu, Liu, and Zhu}{Zhou
  et~al\mbox{.}}{2020}]%
        {zhou2020dast}
\bibfield{author}{\bibinfo{person}{Mingyi Zhou}, \bibinfo{person}{Jing Wu},
  \bibinfo{person}{Yipeng Liu}, \bibinfo{person}{Shuaicheng Liu}, {and}
  \bibinfo{person}{Ce Zhu}.} \bibinfo{year}{2020}\natexlab{}.
\newblock \showarticletitle{DaST: Data-free Substitute Training for Adversarial
  Attacks}. In \bibinfo{booktitle}{\emph{Proceedings of the IEEE/CVF Conference
  on Computer Vision and Pattern Recognition}}. \bibinfo{pages}{234--243}.
\newblock


\end{thebibliography}

\end{document}